\documentclass[a4paper,12pt]{article}
\pdfoutput=1

\usepackage{jheppub}
\usepackage[utf8]{inputenc}
\usepackage[english]{babel}

\usepackage{bm}
\usepackage{amsfonts}
\usepackage{mathbbol}
\usepackage{amsthm}
\usepackage{amssymb}
\usepackage{mathrsfs}
\usepackage{latexsym}
\usepackage{commath} 
\usepackage{float}
\usepackage{subfig}
\usepackage{subcaption}
\usepackage{comment}
\usepackage[makeroom]{cancel}
\usepackage{xcolor}

\definecolor{darkraspberry}{rgb}{0.9,0.,0.3}

\usepackage{soul} 

\begin{document}

\title
{Reflecting Baths and Black Hole Interior Islands in JT Gravity}

\author[a,b]{Dmitry S. Ageev} 
\author[a]{and Anastasia N. Zueva}

\affiliation[a]{Department of Mathematical Methods for Quantum Technologies, Steklov Mathematical
Institute of Russian Academy of Sciences, Gubkina str. 8, Moscow 119991, Russia}
\affiliation[b]{Institute for Theoretical and Mathematical Physics, Lomonosov Moscow State University, 119991 Moscow, Russia}

\emailAdd{ageev@mi-ras.ru}
\emailAdd{zueva@mi-ras.ru}

\abstract{
We study the evolution of entanglement entropy within the island prescription in Jackiw--Teitelboim gravity coupled to finite non-gravitating baths bounded by reflecting  walls. For free massless Dirac matter, we study the evolution of entanglement entropy, explicitly accounting for correlations induced by the reflecting boundaries. For sufficiently short baths, the no-island entropy saturates below the thermodynamic black hole entropy, avoiding the usual information paradox associated with its unbounded growth. For larger baths, reflected correlations drive the exterior quantum extremal surface toward  the future horizon, leading to a temporary disappearance of the the near-horizon exterior island branch. We find that this branch first appears at a scrambling-like scale, and it reappears at a time that is proportional to the size of the bath. We extend the extremization across the horizon using regular Kruskal coordinates and identify real quantum extremal surfaces in the black hole interior. We show that, in the parameter regime studied, interior island configurations provide the dominant contribution to the radiation entropy during the exterior disappearance time interval, resolving the reappearance of the information paradox in an exterior-only description.
}

\maketitle

\clearpage

\section{Introduction}

The black hole information problem asks  how a semiclassical description of Hawking radiation can be reconciled with unitary quantum evolution~\cite{Hawking:1975vcx,Hawking:1976ra}. A useful way to formulate this question is through the fine-grained entropy of the radiation. For a black hole formed in a pure state and evaporating completely, this entropy should eventually decrease and return to zero, following the Page curve~\cite{Page:1993wv}. The semiclassical calculation on a fixed background instead predicts continued entropy growth. A related problem arises for an eternal black hole in equilibrium with a bath: although its thermodynamic properties are stationary, the entropy of radiation collected on the two sides can grow beyond the scale set by the black hole entropy~\cite{Almheiri:2019yqk}. Gravitational calculations of the radiation entropy thus connect the unitarity of black hole evolution with the encoding of interior information in exterior degrees of freedom~\cite{Hartman:2013qma,Maldacena:2015waa, Almheiri:2020cfm}.

The quantum extremal surface (QES) prescription~\cite{Ryu:2006bv,Hubeny:2007xt,Faulkner:2013ana,Engelhardt:2014gca, Akers:2019lzs} and the island formula provide a framework for addressing this question. Early calculations of evaporating black holes found a transition in the dominant quantum extremal surface that reproduces the expected black hole Page curve~\cite{Penington:2019npb,Almheiri:2019psf}. The corresponding prescription for the radiation entropy was formulated using a higher-dimensional holographic description~\cite{Almheiri:2019hni} and subsequently derived from replica wormholes in the gravitational path integral~\cite{Almheiri:2019qdq,Penington:2019kki}. For a radiation region $\mathcal R$ in a non-gravitating bath, it takes the form
\begin{equation}
 S(\mathcal R)=\min_{\mathcal I}\operatorname*{ext}_{\partial\mathcal I}
 \left[
 S_{\mathrm{grav}}(\partial\mathcal I)
 +S_{\mathrm{matter}}(\mathcal I\cup\mathcal R)
 \right],
 \label{eq:intro-island}
\end{equation}
where the minimization includes the trivial configuration without an island. The first term is the gravitational entropy associated with the island boundary, while the second is the entropy of quantum fields on the union of the island and the radiation region. The behavior of radiation entropy near the endpoint of black hole evaporation, including its dependence on the assumed evolution of mass and charge, has also been investigated in~\cite{Arefeva:2022guf}. The existence and dominance of entanglement islands, together with the associated radiation entropy, can depend sensitively on the matter state, the size of the entangling region, and the infrared regularization \cite{Ageev:2022hqc,Ageev:2022qxv,Ageev:2023mzu}.

Jackiw--Teitelboim (JT) gravity~\cite{Jackiw:1984je,Teitelboim:1983ux} coupled to a conformal field theory (CFT) provides a particularly convenient framework for this analysis. The metric has fixed local negative curvature, whereas the dilaton determines the gravitational entropy and responds to the matter stress tensor. Transparent coupling to a non-gravitating CFT bath makes it possible to study both evaporation and equilibrium within this framework.  In particular, the evaporating JT model of Ref.~\cite{Almheiri:2019psf} relates the motion of the quantum extremal surface to information recovery on a scrambling time scale. The real time replica construction of Ref.~\cite{Goto:2020wnk} establishes the island rule for evaporating black holes with Lorentzian interior islands. These results motivate extending the search for quantum extremal surfaces across the horizon and into the black hole interior.

An important question is how this picture changes when the radiation cannot escape into an indefinitely large reservoir. A natural example is provided by asymptotically anti-de Sitter black holes with reflecting boundary conditions at conformal infinity, which return Hawking radiation toward the black hole~\cite{Hawking:1982dh,Witten:1998zw,Maldacena:2001kr}. Another possibility is to introduce reflecting walls explicitly, placing the black hole in a finite cavity~\cite{York:1986it}. A reflecting wall introduces a finite propagation scale and returns correlations toward the black hole. It can therefore change the entropy competition even when the local geometry remains stationary. In a higher-dimensional setting, studies of a two-sided Schwarzschild black hole in a reflecting cavity found both saturation of the no-island entropy and a temporary disappearance of an island solution, termed a blinking island~\cite{Ageev:2023hxe,Ageev:2021ipd}. The influence of reflecting boundaries on island formation has also been studied in fuzzball-inspired models, where a reflecting stretched horizon replaces the black hole horizon~\cite{Ageev:2026ezm}. Studies of finite radiation intervals in JT gravity have identified islands that dominate the entropy only for a finite time, together with quantum extremal surfaces that move across the horizon~\cite{Hollowood:2021wkw, Hollowood:2021nlo,Gyongyosi:2022vaf}. Transitions between interior and exterior QES configurations also occur when evaporation is induced by a shockwave~\cite{Hollowood:2020cou,Craps:2022ahp}. These results suggest that the loss of an exterior solution may be accompanied by a transition to another QES branch, allowing an island to continue contributing to the radiation entropy.

Reflecting boundaries also raise the question of how the return time of Hawking radiation compares with the black hole scrambling time. In AdS, outgoing radiation can reach the conformal boundary and return to the black hole, introducing a propagation scale that depends on the geometry and the emission radius~\cite{Ling:2021nad}. Comparing this scale with the scrambling time requires a common time coordinate and a consistent treatment of the near-horizon redshift, whose logarithmic delay is closely related to scrambling~\cite{Shenker:2013pqa,Langhoff:2020jqa,Barbon:2011pn,Asplund:2015eha}. If reflected radiation returns before scrambling is complete, its correlations must be retained when describing information exchange between the black hole and its surroundings. This motivates studying how boundary reflections modify radiation entropy and the quantum extremal surfaces that determine it.

In this paper, we study a two-sided eternal JT black hole coupled transparently to two identical flat, non-gravitating baths, each limited by a perfectly reflecting conformal wall. The walls lie in the baths, leaving a well-defined non-gravitating radiation subsystem. We consider the matter sector to consist of $n$ copies of free massless Dirac fermions, with perfectly reflecting boundary conditions (BCFT), and prepare the two-sided Hartle--Hawking state. Radiation is collected in the two segments between a bath endpoint and the corresponding wall. This setup allows us to retain the reflected correlations explicitly, derive the generalized entropy, and follow symmetric quantum extremal surfaces through the horizon. Our aim is to determine when the finite size of the bath is sufficient to avoid the Page formulation of the information paradox and when its resolution requires quantum extremal surfaces inside the horizon.

$\,$

Now let us briefly summarize our findings. First we obtain exact free-fermion expressions for the no-island and symmetric island entropies, including the boundary image terms. The no-island entropy grows initially and then saturates at a value controlled by the bath size.  When the bath is sufficiently short, the no-island entropy saturates below the thermodynamic black hole entropy, so that the usual information paradox associated with its growth
does not arise. However, this condition does not necessarily imply the dominance of the no-island configuration. The physical entropy is determined by comparing the complete generalized entropies, and the corresponding Page transition depends on the competition between the bath propagation scale $L-b$ and the Page time $t_{\rm Page} \simeq 3S_{\rm BH}/c+b+\log 2$. For larger baths, nontrivial island configurations can become essential for determining the radiation entropy.

The duration of the exterior blink has a useful separation of scales. In the distant wall regime in which the early branch exists the blink time is given by a scrambling-like  resembling form \cite{Hayden:2007cs,Sekino:2008he,Almheiri:2019psf}, namely
\begin{equation}
 t_{\mathrm{dis}}
 \simeq \frac{\beta}{2\pi}(L-b),\qquad
 \Delta t_{\mathrm{blink}}
 \simeq \frac{3\beta}{2\pi} b+\frac{3\beta}{4\pi}
 \log\!\left[\frac6c\left(S_{\mathrm{BH}}-2S_0\right)\right].
\end{equation}
The wall position fixes the onset of the exterior blink, while its leading duration is independent of the bath size. The logarithmic dependence arises when the exponentially decreasing image contributions become comparable to the small terms that determine the near-horizon QES. This suggests that recovery of the exterior branch probes a reorganization of correlations on a near-horizon information-processing scale. 

To determine what replaces the missing exterior branch, we formulate the generalized entropy in Kruskal coordinates that remain regular at the future horizon. We then extend the extremization into the black hole interior, derive a controlled approximate interior solution, and construct real interior extrema numerically. Our analysis shows that interior QES branches can sustain the dominant island contribution after the Page transition, including throughout the interval in which the exterior branch is absent. The apparent return of the information paradox in an exterior-only calculation is thus avoided within the studied symmetric sector. This indicates that nontrivial bath boundary conditions can make the near-horizon interior essential to the generalized entropy calculation. 

Finally, we calculate the semiclassical backreaction (see \cite{Pedraza:2021cvx}) and find that in the Hartle--Hawking state it reduces to a constant shift of the dilaton. The QES equations remain unchanged, although the entropy shift can modify the Page transition. Since the local stress tensor has no explicit dependence on the wall position, the reflected correlations affect the QES through the matter entropy.

The paper is organized as follows. Section~\ref{sec:JTandBath} introduces the finite bath geometry and computes the no-island entropy. Section~\ref{sec:Islands} derives and analyzes the symmetric exterior QES equations, including the critical bath size and the scrambling-like blinking interval. Section~\ref{iqes:section} constructs interior extrema and compares the entropy branches. Section~\ref{sec:backreaction} studies the matter stress tensor and dilaton backreaction, and Section~\ref{sec:Conclusion} presents our conclusions.

\section{JT gravity coupled to a finite bath}\label{sec:JTandBath}

In this section we set up the model and derive the entropy of the configuration with the trivial island. We consider a two-sided eternal Jackiw--Teitelboim black hole coupled transparently to two identical non-gravitating flat baths~\cite{Almheiri:2019yqk,Almheiri:2019qdq}, each of which ends at a reflecting wall. The crucial point is that the wall lies in the flat bath rather than in the gravitating AdS$_2$ region. The standard separation between the gravitating region and the non-gravitating radiation system is therefore preserved, but the radiation system now has a finite infrared scale~\cite{Kruthoff:2021vgv,Hollowood:2021lsw}. We first describe the geometry and conventions, then collect the BCFT entropy formula in the presence of the wall, and finally continue to Lorentzian signature to obtain the generalized entanglement entropy without an island.

For explicit calculations we take the matter sector to be a free massless Dirac theory of total central charge $c$. For free fermions, multi-interval entropies with a reflecting boundary can be expressed explicitly in terms of direct and image distances~\cite{Casini:2005rm,Rottoli:2022plr,Hollowood:2021wkw}. In contrast, for a generic interacting CFT, the corresponding two-interval entropy would involve an additional non-universal function of the cross-ratio~\cite{Calabrese:2009ez,Mintchev:2020uom}. Therefore, the exact finite time formulas given below should be understood as specific to this solvable matter sector. However, the late time analysis near the horizon of the QES is more robust and applicable to more general cases.

\subsection{Geometry and conventions}
Jackiw--Teitelboim (JT) gravity provides a simple framework for studying black hole physics in two dimensions. The spacetime geometry has constant negative curvature, while the dynamical dilaton determines the gravitational contribution to the black hole entropy. The Lorentzian JT action takes the form~\cite{Jackiw:1984je,Teitelboim:1983ux}
\begin{equation}\label{eq:ClassicalJT}
    \begin{aligned}
        I_\text{JT}&= I_\text{JT}^{bulk} + I_\text{JT}^{bdy},\\
        I_\text{JT}^{bulk} &=\frac{1}{16\pi G}\bigg[\int_M d^2 x\sqrt{-g}\,\Phi_0\,R +\int_M d^2 x\sqrt{-g}\,\Phi\left(R+2\right)\bigg], \\
        I_\text{JT}^{bdy} & = \frac{1}{8\pi G}\bigg[\int_{\partial M} dt \sqrt{-\gamma}\, \Phi_0\,K + \int_{\partial M} dt\sqrt{-\gamma}\,\Phi\left(K-1\right)\bigg],
    \end{aligned}
\end{equation}
where $G$ is two-dimensional Newton's constant, $K$ is extrinsic curvature of $\partial M$ with induced metric $\gamma$, $\Phi$ is dilaton and $\Phi_0$ is a constant proportional to the entropy of the higher-dimensional black hole geometry.

The dilaton equation of motion fixes the Ricci scalar to $R=-2$, so the spacetime is locally AdS$_2$. In the black hole patch, the metric takes the form
\begin{equation}\label{eq:BHpatch}
    ds^2=-(r^2-r_h^2)dt^2+\frac{dr^2}{r^2-r_h^2},
    \qquad r_h=\frac{2\pi}{\beta},
\end{equation}
where $r_h$ is the horizon radius and $\beta$ is the inverse Hawking temperature. Introducing the radial coordinate $\sigma_{\rm phys}$ through the transformation
\begin{equation}
    r=-r_h\coth\left(\frac{2\pi\sigma_{\rm phys}}{\beta}\right),
\end{equation}
and denoting the corresponding physical time coordinate by $t_{\rm phys}$, we can rewrite the metric \eqref{eq:BHpatch} in conformally flat form,
\begin{equation}\label{eq:MetricNullLor}
    ds^2=\frac{4\pi^2}{\beta^2}
    \frac{-dt_{\rm phys}^2+d\sigma_{\rm phys}^2}
    {\sinh^2\left(\frac{2\pi\sigma_{\rm phys}}{\beta}\right)}.
\end{equation}
In these coordinates, the AdS$_2$ boundary at $r\to\infty$ corresponds to $\sigma_{\rm phys}\to0^-$, while the horizon is approached as $\sigma_{\rm phys}\to-\infty$.

\begin{figure}[t!]
    \centering
    \includegraphics[width=0.7\linewidth]{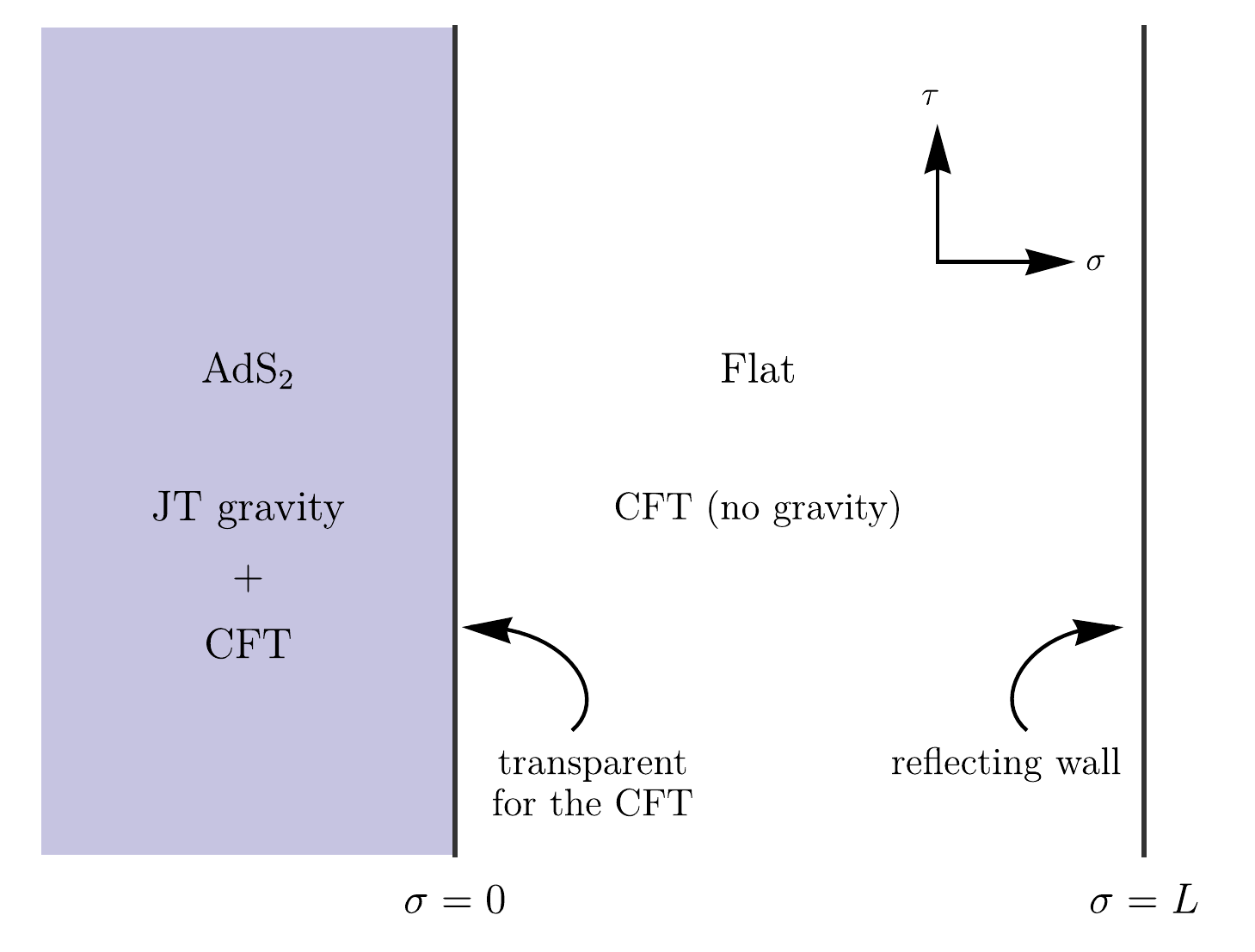}
    \caption{AdS$_2$ gravitating region glued to the flat bath through a transparent interface at $\sigma=0$, with a reflecting wall imposed at $\sigma=L$.}
    \label{fig:TauSigmaRegion}
\end{figure}

To study the effects of a finite radiation reservoir, we couple the gravitating AdS$_2$ region, occupying $\sigma_{\rm phys}<0$, to a flat, non-gravitating bath extending over $0<\sigma_{\rm phys}<L_{\rm phys}$, as illustrated in Fig.~\ref{fig:TauSigmaRegion}. The interface at $\sigma_{\rm phys}=0$ is transparent to the matter fields, allowing them to propagate between the gravitating region and the bath. At the outer endpoint $\sigma_{\rm phys}=L_{\rm phys}$, we impose a reflecting conformal boundary condition, which introduces a finite length scale and modifies the correlations of the matter fields. Since gravity is dynamical only in the AdS$_2$ region, the radiation subsystem in the bath retains the usual non-gravitating Hilbert-space interpretation. The regularization and matching conditions at the AdS$_2$--bath interface are discussed in Section~\ref{sec:backreaction}.

Moving on to the Euclidean signature with $t_{\rm phys} = i\tau_{\rm phys}$, the metric in complex coordinates $y$ for a black hole and a flat space glued to the both sides of it takes the form
\begin{gather}\label{eq:MetricY}
ds^2_\text{JT}=\frac{4\pi^2}{\beta^2}\frac{d y_{\rm phys} d\bar y_{\rm phys}}{\sinh^2\frac{\pi}{\beta}(y_{\rm phys}+\bar y_{\rm phys})}, \quad ds^2_\text{bath}=
dy_{\rm phys} d\bar y_{\rm phys}, \\
y_{\rm phys} = \sigma_{\rm phys} + i \tau_{\rm phys},\quad \bar y_{\rm phys} = \sigma_{\rm phys} - i\tau_{\rm phys}, \quad \tau_{\rm phys} \sim \tau_{\rm phys} + \beta
\end{gather}
For the calculations below, it is convenient to use dimensionless thermal coordinates adapted to the eternal black hole solution. Denoting the inverse temperature by $\beta$, we introduce the scale
\begin{align}
\kappa &= \frac{2\pi}{\beta} \, .
\end{align}
In what follows, we use dimensionless coordinates defined by \mbox{$\sigma = \kappa \sigma_{\rm phys}$} and \mbox{$\tau = \kappa \tau_{\rm phys}$}. The Euclidean time coordinate $\tau$ then has period $2\pi$.

Under the exponential map
\begin{align}
w &= e^y \, , \qquad \bar w = e^{\bar y} \, ,
\end{align}
the Euclidean thermal strip becomes a disk. The gravitating region maps to the unit disk,
\begin{align}
\sigma<0
\qquad &\Longleftrightarrow \qquad
|w|<1 \, ,
\end{align}
while the bath maps to the annulus
\begin{align}
0<\sigma<L
\qquad &\Longleftrightarrow \qquad
1<|w|<R \, , \qquad R=e^L \, .
\end{align}
Thus $|w|=1$ is the transparent AdS$_2$--bath interface and $|w|=R$ is the physical reflecting wall. From the point of view of the matter path integral, the full geometry is the disk $|w|<R$ with a BCFT boundary only at $|w|=R$.

In the $w$ coordinate the JT region carries the Poincar\'e disk metric,
\begin{align}
ds^2_{\rm JT}
&=
\frac{4 \, dw \, d\bar w}{(1-w\bar w)^2} \, ,
\end{align}
while the bath metric is flat,
\begin{align}
ds^2_{\rm bath}
&=
\frac{1}{\kappa^2}
\frac{dw \, d\bar w}{w\bar w} \, .
\end{align}
It is convenient to write both metrics as
\begin{align}
ds^2 &= \Omega(w,\bar w)^{-2} \, dw \, d\bar w \, ,
\end{align}
with Weyl factors
\begin{equation}
\Omega_{\rm JT}(w,\bar w) = \frac{1-w\bar w}{2},\quad
\Omega_{\rm bath}(w,\bar w) = \kappa |w|.
\end{equation}

The equation of motion for the metric give an expression for the dilaton stress-energy tensor
\begin{equation}\label{eq:StressTensorDilaton}
    T_{\mu\nu}^\Phi = 0, \quad T_{\mu\nu}^\Phi \equiv -\frac{2}{\sqrt{-g}}\frac{\delta I_\text{JT}}{\delta g^{\mu\nu}} = -\frac{1}{8\pi G}\left(g_{\mu\nu}\nabla^2 -\nabla_\mu\nabla_\nu - g_{\mu\nu}\right)\Phi.
\end{equation}
Therefore the solution for the dilaton, which is  defined only in the JT region, is~\cite{Maldacena:2016upp}
\begin{align}\label{eq:Dilaton}
\Phi(w,\bar w) &= \Phi_h \, \frac{1+w\bar w}{1-w\bar w} \, ,
\end{align}
where $\Phi_h$ is the horizon value. Finally, we assume that the system is in the semiclassical regime,
characterized by
\begin{equation}\label{eq:alpha}
    \alpha \equiv \frac{cG}{3\Phi_h} \ll 1,
\end{equation}
and neglect the backreaction of the matter stress tensor. Accordingly, we evaluate the CFT contribution to the generalized entropy on the unperturbed classical JT background, with the dilaton profile given by \eqref{eq:Dilaton}.

\subsection{Entanglement entropy for the single region on a disc}\label{sec:SingleInterval}

In order to calculate the entanglement entropy in a system that consists of a black hole and a finite-sized bath, we consider a conformal field theory with a boundary (BCFT). Namely, we consider two-dimensional massless Dirac fermions with perfectly reflecting boundary conditions and a total central charge $c$. Then, for $n$ intervals
\begin{equation}
    \mathcal{R}=[z_{a_1},z_{b_1}]\cup ... \cup [z_{a_m},z_{b_m}]
\end{equation}
their entanglement entropy has the following form \cite{CARDY1984514, Cardy:1986gw, Cardy:1989ir,Mintchev:2020uom, Rottoli:2022plr, Kruthoff:2021vgv,Casini:2005rm,Reyes:2021npy} 
\begin{equation}\label{eq:BCFTent}
    \begin{aligned}
        S_\text{CFT}(\mathcal{R})=&
        \frac{c}{3}\sum_{i,j=1}^n\ln |z_{a_i}-z_{b_j}|
        -\frac{c}{3}\sum_{i<j}^n \ln |z_{a_i}-z_{a_j}| |z_{b_i}-z_{b_j}|
        -\frac{c}{3}\,n \ln \varepsilon \\
        &+
        \frac{c}{6}\sum_{i,j=1}^n
        \ln |z_{a_i}-\bar{z}_{a_j}| |z_{b_i}-\bar{z}_{b_j}|
        - \frac{c}{6} \sum_{i,j=1}^n
        \ln |z_{a_i}-\bar{z}_{b_j}| |z_{b_i}-\bar{z}_{a_j}|.
    \end{aligned}
\end{equation}
For the single interval $[z_1, z_2]$, the entanglement entropy on a half-plane reduces to 
\begin{equation}\label{eq:EntCFTz}
    S_\text{CFT}(z_1,z_2)=\frac{c}{6}\ln \left(\frac{|z_1-z_2|^2|z_1-\bar z_1||z_2-\bar z_2|}{\epsilon_1^{(z)}\epsilon_2^{(z)}|z_1-\bar z_2||z_2-\bar z_1|}\right).
\end{equation}
Here $z_i-z_j$ denotes the coordinate separation between insertions, whereas $z_i-\bar z_j$ denotes the distance between the insertion at $z_i$ and the image of the insertion at $z_j$ across the reflecting boundary.

To apply the BCFT formula \eqref{eq:EntCFTz} we map the disk $|w|<R$ to the upper half-plane using the following transformations
\begin{align}\label{eq:MapToDisk}
z &= i \, \frac{R+w}{R-w}, \quad \bar z
= - i \, \frac{R+\bar w}{R-\bar w}.
\end{align}
Then, the distances in the expression for entropy, in coordinates on the disk, takes the form
\begin{equation}\label{eq:ZCombinations}
 z_i-z_j = \frac{2iR(w_i-w_j)}{(R-w_i)(R-w_j)}, \qquad
z_i-\bar z_j = \frac{2i(R^2 - w_i\bar w_j)}{(R-w_i)(R-\bar w_j)},
\end{equation}
and  UV cutoff at a point $w_i$ needs to be multiplied by the local Weyl factor, leading to
\begin{equation}\label{eq:cutoffs}
    \epsilon^{(z)}=\bigg|\frac{dz}{dw}\bigg|\Omega \epsilon^{(w)}, \quad \frac{dz}{dw} = \frac{2iR}{(R-w)^2}.
\end{equation}

Therefore, by substituting the corresponding combinations of \eqref{eq:ZCombinations} and cutoffs \eqref{eq:cutoffs} in the expression \eqref{eq:EntCFTz}, for the single interval with endpoints $w_1$ and $w_2$, the entanglement entropy on the disk takes the form
\begin{align}\label{eq:EntSingleRegion}
S_\text{CFT}(w_1,w_2) &= \frac{c}{6} \log \left[ \frac{
(w_1-w_2)(\bar w_1-\bar w_2) }{ \epsilon_1^{(w)} \epsilon_2^{(w)} \, \Omega_1 \Omega_2 } \frac{
(R^2-w_1\bar w_1)(R^2-w_2\bar w_2) }{ (R^2-w_1\bar w_2)(R^2-w_2\bar w_1) } \right] + S_{\rm bdy} \, .
\end{align}
The constant $S_{\rm bdy}$ is the Affleck--Ludwig boundary entropy \cite{Affleck:1991tk} together with possible scheme-dependent terms. We will usually suppress it in comparisons because it does not affect extremization with respect to island endpoints, although it can shift Page times by terms of order one.

\subsection{Entanglement entropy with trivial island}

\begin{figure}[t!]
    \centering
    \includegraphics[width=0.7\linewidth]{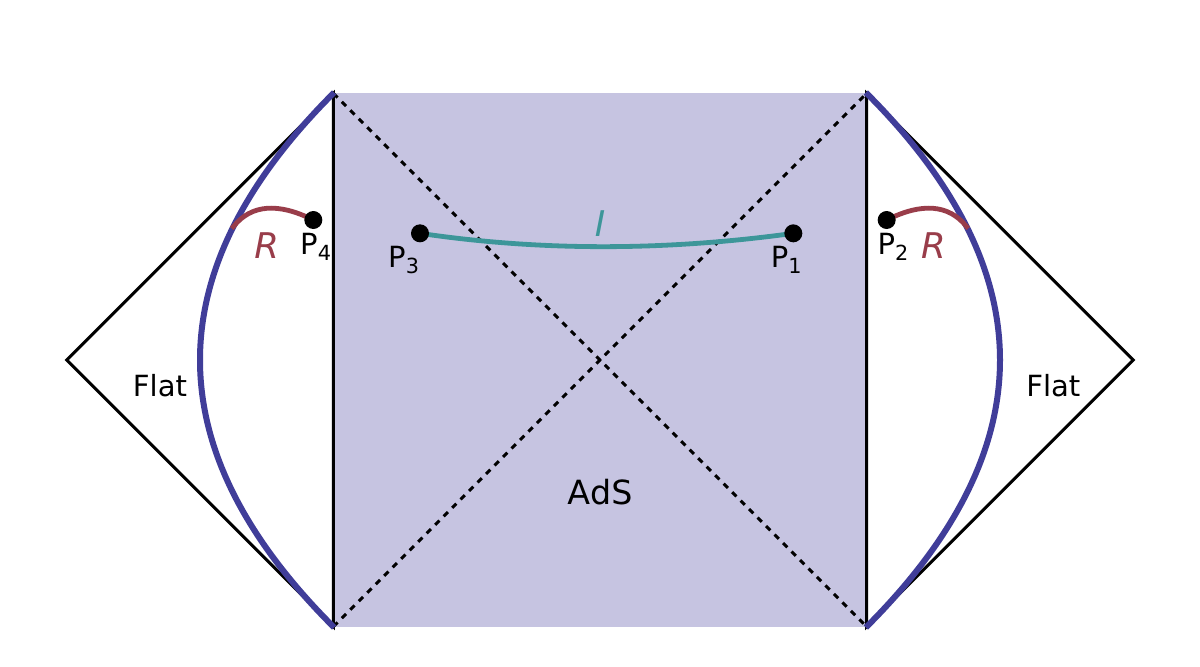}
    \caption{Eternal AdS$_2$ black hole, glued to Minkowski space on both sides, truncated by a reflecting wall. Region $I$ corresponds to the island, Hawking radiation is collected in region $R$.}
    \label{fig:IslandRegion}
\end{figure}
We now calculate the entanglement entropy of the radiation region in the absence of an island. The radiation region consists of two bath segments, each extending from an endpoint in the flat bath to a reflecting wall (see Fig. \ref{fig:IslandRegion}),
\begin{align}
\mathcal{R}=[P_4,{\rm wall}_{\rm left}]\cup [P_2,\text{wall}_{\rm right}],
\end{align}
where the left endpoint at the same exterior time is obtained by the standard thermofield-double shift by half of the Euclidean thermal circle.
The entropy of $\mathcal{R}$ in the no-island saddle is the single-interval entropy of the complement $[P_4, P_2]$. After continuing to Lorentzian signature, we parametrize the endpoints as
\begin{equation}
    P_2 = (b,t_b), \quad P_4 = (b,-t_b+i\pi),
\end{equation}
with 
\begin{equation}\label{eq:RadiationEndpoints}
    \begin{aligned}
        w_2 &= e^{b+t_b},\\
        w_4 &= -e^{b-t_b},
    \end{aligned} \qquad 
    \begin{aligned}
        \bar w_2 &= e^{b-t_b},\\
        \bar w_4 &= -e^{b+t_b} .
    \end{aligned}
\end{equation}
Let us consider in more detail the contributions that are included in entropy \eqref{eq:EntSingleRegion}. Writing $B=e^b$, we obtain the following wall-independent term
\begin{equation}\label{eq:temp1}
(w_4-w_2)(\bar w_4-\bar w_2) = 4B^2\cosh^2 t_b, \quad \Omega_2\Omega_4 = \kappa^2B^2 \, .
\end{equation}
The contribution from the boundary is controlled by the bath distance from the endpoint to the wall via following combinations
\begin{equation}\label{eq:temp2}
    \begin{gathered}
        R^2-w_2\bar w_2 =R^2-w_4\bar w_4= 2RB\sinh (L-b), \\
        (R^2-w_2\bar w_4)(R^2-w_4\bar w_2) = 2R^2B^2 \left( \cosh2(L-b)+\cosh2t_b \right).
    \end{gathered}
\end{equation}
Substituting \eqref{eq:temp1} and \eqref{eq:temp2} into the formula \eqref{eq:EntSingleRegion} for the matter part of the entropy on the disk gives
\begin{align}\label{eq:EntNoIsl}
S_\text{noIsl}[P_4,P_2] &= \frac{c}{6} \log \left[
\frac{4\cosh^2 t_b}{(\kappa\epsilon_b)^2}
\frac{2\sinh^2(L-b)}{\cosh2(L-b)+\cosh2t_b}
\right]\, .
\end{align}

This expression makes the role of the finite bath transparent. The first factor is the usual thermal growth of the two-sided black hole \cite{Hartman:2013qma}. The second factor is the BCFT image contribution from the reflecting wall. For $t_b \ll L-b$ the image term is close to one, so the entropy agrees with the infinite bath result. When $t_b \sim L-b$ the reflected image becomes important, and at later times it suppresses  the growth. In the late time regime \(t_b\gg L-b\), the time dependence cancels and one obtains the finite saturation value
\begin{align}\label{eq:EntSat}
S_{\rm noIsl}^{\infty} = \frac{c}{3}\log\left[\frac{2\sinh (L-b)}{\kappa\epsilon_b}\right] .
\end{align}
The competition between the contribution of this trivial island and the island saddle is crucial to the further analysis.

This entropy provides a complementary way to understand the role of the finite bath. By differentiating the exact result, we obtain
\begin{equation}
    \frac{dS_{\rm noIsl}}{dt_b} = \frac{2c}{3}
    \frac{\tanh t_b\,\sinh^2 (L-b) }{ \cosh 2d+\cosh 2t_b}>0,
\end{equation}
so the entropy approaches its finite saturation value monotonically.
Moreover, for \mbox{$L-b\gg1$},
\begin{equation}
    S_{\rm noIsl}^{\infty} =\frac{c}{3}(L-b) +\text{UV and boundary terms}
    +\mathcal O(e^{-2(L-b)}).
\end{equation}
Since the thermal entropy density of a two-dimensional CFT is~\cite{Calabrese:2004eu}
\begin{equation}
    s_{\rm th}=\frac{\pi c}{3\beta}=\frac{c\kappa}{6},
\end{equation}
the leading part is precisely
\begin{equation}
    \frac{c}{3}(L-b) = 2s_{\rm th} \frac{(L-b)}{\kappa} .
\end{equation}
The saturation of the no-island saddle can therefore be interpreted as the finite thermal entropy density of the two bath segments.

\section{Island saddle for the symmetric region}\label{sec:Islands}

\subsection{Entanglement entropy with island}

We now extend the calculation of entanglement entropy to configurations with nontrivial islands. According to the island prescription~\cite{Engelhardt:2014gca,Almheiri:2019hni}, the entropy of a radiation region $\mathcal R$ is obtained by extremizing the generalized entropy with respect to the island boundaries and selecting the minimum among the admissible extrema, including the configuration without an island,
\begin{equation}
S(\mathcal R)=\min_{\mathcal I}\left\{\underset{\partial\mathcal I}{\rm ext}\;
S_{\rm gen}(\mathcal I \cup \mathcal{R})\right\},
\end{equation}
where the generalized entropy combines gravitational and matter contributions,
\begin{equation}
S_{\rm gen}(\mathcal I \cup \mathcal{R})=
\frac{\mathrm{Area}(\partial\mathcal I)}{4G} +
S_{\rm CFT}(\mathcal I \cup \mathcal{R}).
\end{equation}
In two-dimensional JT gravity, the gravitational contribution is determined by the dilaton evaluated at the island endpoints. For the two-endpoint island configuration considered below, the generalized entropy therefore takes the form
\begin{equation}
S_{\rm gen}=2S_0+\frac{\Phi(\partial\mathcal I)}{4G}
+S_{\rm CFT}(\mathcal I\cup\mathcal R),
\label{eq:JTGenEnt}
\end{equation}
where $\Phi(\partial\mathcal I)$ denotes the sum of the dilaton values at the two island endpoints, and $S_0$ is the topological contribution to the black hole entropy. The extremization of \eqref{eq:JTGenEnt} determines the quantum extremal surfaces, while the comparison of their generalized entropies with the no-island contribution determines the physical radiation entropy and possible Page transitions. To evaluate the matter contribution for a nontrivial island, we generalize the single-interval BCFT calculation of Section~\ref{sec:SingleInterval} to the union of two intervals.

We consider a left-right symmetric island whose endpoints are located at the same radial position $\sigma=-a$ in the two gravitating regions, as illustrated in Fig.~\ref{fig:IslandRegion}. As in the no-island configuration, the radiation region consists of two bath segments extending from the endpoints $P_2$ and $P_4$ to the corresponding reflecting walls,
\begin{equation}
    \mathcal{R}=[P_4,{\rm wall}_{\rm left}]\cup [P_2,{\rm wall}_{\rm right}],
\end{equation}
while the island is the interval connecting the two points $P_3$ and $P_1$,
\begin{equation}
    \mathcal{I}=[P_3,P_1].
\end{equation}
In dimensionless thermal coordinates, the four endpoints are parametrized as
\begin{align}
    P_1=(-a,t_a),\qquad
    P_2=(b,t_b),\qquad
    P_3=(-a,-t_a+i\pi),\qquad
    P_4=(b,-t_b+i\pi),
\end{align}
where the coordinates of the left endpoints are obtained through the standard thermofield-double continuation. Under the exponential map to the disk, the island endpoints take the form
\begin{equation}
\begin{aligned}
    w_1&=e^{-a+t_a},\\
    w_3&=-e^{-a-t_a},
\end{aligned}
\qquad
\begin{aligned}
    \bar w_1&=e^{-a-t_a},\\
    \bar w_3&=-e^{-a+t_a}.
\end{aligned}
\end{equation}
The corresponding disk coordinates of the radiation endpoints $P_2$ and $P_4$ are given by Eq.~\eqref{eq:RadiationEndpoints}.

For the symmetrical island configuration, the gravitational contribution to the generalized entropy \eqref{eq:JTGenEnt} is determined by the dilaton profile \eqref{eq:Dilaton} as follows
\begin{equation}\label{eq:GravContrib}
    S_\text{grav}=\frac{1}{4 G}\bigg(2\Phi_0+ \Phi(w_1,\bar w_1)+\Phi(w_3,\bar w_3)\bigg)=2 S_0+\frac{\Phi_h}{2G}\coth a, \quad S_0=\frac{\Phi_0}{4 G}.
\end{equation}
Since the full two-sided state is pure, the matter entropy of \(\mathcal{I}\cup\mathcal{R}\) is equal to the matter entropy of its complement, which is the union of the two intervals
\begin{align}
[P_4,P_3]\cup [P_1,P_2] \, .
\end{align}
As in the previous section, we calculate the entanglement entropy of the union of these two segments using  formula \eqref{eq:BCFTent}. The expression for the matter part of the entropy can be naturally divided into a wall-independent term and a term that depends on the wall position
\begin{equation}
    S_\text{CFT}(\mathcal{I}\cup\mathcal{R}) =S_\text{CFT}^{(0)} + S_\text{CFT}^{\rm wall}.
\end{equation}
After absorbing the ultraviolet divergence at the gravitational endpoints into the renormalized JT entropy $S_0$, the wall-independent matter contribution takes the compact form
\begin{align}\label{eq:exteriorIslandCFT0}
S_\text{CFT}^{(0)} = \frac{c}{3} \log\left[
\frac{ 4\cosh t_a\,\cosh t_b }{ \kappa\epsilon_b\,\sinh a } \,
\frac{ \cosh(a+b)-\cosh(t_a-t_b) }{ \cosh(a+b)+\cosh(t_a+t_b) } \right] .
\end{align}
The ``wall contribution'' is derived from the corresponding image factors, and the wall part of the matter entropy becomes
\begin{align}
S_\text{CFT}^{\rm wall}&= \frac{c}{3} \log\left[
\frac{ \cosh(2L+a-b)+\cosh(t_a+t_b) }{ \cosh(2L+a-b)-\cosh(t_a-t_b) } \right] \notag\\
&\qquad + \frac{c}{6} \log\left[ \frac{ 4\sinh^2(L+a)\sinh^2(L-b) }{ \left(\cosh 2(L+a)+\cosh 2t_a\right)\left(\cosh 2(L-b)+\cosh 2t_b\right) } \right] .
\end{align}
We keep the bath cutoff explicit and denote it by $\epsilon_b$, while cutoffs at QES endpoints in the gravitating region are absorbed into the renormalized quantities $S_0$ and $\Phi/G$.

Therefore, the generalized entropy for the symmetric island  configuration, up to renormalized and additive boundary terms, takes the form
\begin{align}\label{eq:EntGen}
S_{\rm gen} = 2S_0+\frac{\Phi_h}{2G}\coth a + S_\text{CFT}^{(0)} + S_\text{CFT}^{\rm wall}\,.
\end{align}
This formula is exact for free Dirac fermions in the finite bath with a reflecting wall, and it reduces to the standard eternal JT plus bath expression~\cite{Almheiri:2019yqk,Almheiri:2019qdq} when $L\to\infty$.

\subsection{QES solutions}
The time of appearance and position of the island are determined by solving a set of equations on a quantum extremal surface (QES). The QES equations are obtained by extremizing the generalized entanglement entropy \eqref{eq:EntGen} with respect to $a$ and $t_a$
\begin{equation}
    \begin{cases}
        \begin{aligned}
            \partial_{t_a} S_\text{gen}&=0,\\
            \partial_{a} S_\text{gen}&=0.
        \end{aligned}
    \end{cases}
\end{equation}
The expression for the time derivative has the form
\begin{equation}\label{eq:QESTime}
\begin{aligned}
    \partial_{t_a}S_{\rm gen} &= \frac{c}{3} \Bigg[ \tanh t_a
-\frac{\sinh(t_a-t_b)}{\cosh(a+b)-\cosh(t_a-t_b)}
-\frac{\sinh(t_a+t_b)}{\cosh(a+b)+\cosh(t_a+t_b)}
\\ 
&+\frac{\sinh(t_a+t_b)}{\cosh(2L+a-b)+\cosh(t_a+t_b)} +\frac{\sinh(t_a-t_b)}{\cosh(2L+a-b)-\cosh(t_a-t_b)} 
\\
& -\frac{\sinh 2t_a}{\cosh 2(L+a)+\cosh 2t_a} \Bigg] .
\end{aligned}
\end{equation}
while the radial derivative can be written as
\begin{align}\label{eq:QESRadial}
\partial_a S_{\rm gen}&=
-\frac{\Phi_h}{2G}\frac{1}{\sinh^2 a} + \frac{c}{3} \Bigg[\coth(L+a) -\frac{\sinh 2(L+a)}{\cosh 2(L+a)+\cosh 2t_a}
\notag \\
&  -\coth a
+\frac{\sinh(a+b)}{\cosh(a+b)-\cosh(t_a-t_b)}
-\frac{\sinh(a+b)}{\cosh(a+b)+\cosh(t_a+t_b)}
\notag\\
& + \frac{\sinh(2L+a-b)}{\cosh(2L+a-b)+\cosh(t_a+t_b)} - \frac{\sinh(2L+a-b)}{\cosh(2L+a-b)-\cosh(t_a-t_b)} \Bigg] .
\end{align}
Before solving these equations numerically, let us analyze them in a semiclassical setting, $\alpha\ll1$, where it is natural to assume that the island is located close to the horizon, so
\begin{equation}
    a\gg 1, \quad \text{and} \quad e^{-a}=\mathcal{O}(\alpha).
\end{equation}
In addition, we assume that all hyperbolic-function arguments appearing in Eqs.~\eqref{eq:QESTime} and \eqref{eq:QESRadial} are parametrically large,
\begin{equation}
    \begin{gathered}
        a+b,\quad 2L+a-b,\quad L+a,\quad t_{a,b}\;\gg 1,\\
        a+b-|t_a-t_b|\gg 1,\quad 2L+a-b-|t_a-t_b|\gg 1,
    \end{gathered}
\end{equation}
allowing us to expand the hyperbolic functions in their large argument limit and retain only terms that contribute to the QES equations at leading order. Terms that are exponentially suppressed by these large parameters are consistently neglected. Under these assumptions, the QES equations reduce to
\begin{equation}\label{eq:QESeqFull}
    \begin{aligned}
        &e^{-(a+b)}\left(1-e^{-2(L-b)}\right)(e^{-(t_a-t_b)} -e^{t_a-t_b})+\frac{1}{1+\exp{[(t_a+t_b)-(a+b)}]}+\\
        &\quad + \frac{\exp{[(t_a+t_b)-(2L+a-b)}]}{1+\exp{[(t_a+t_b)-(2L+a-b)}]}-\frac{\exp{[2t_a-2(L+a)}]}{1+\exp{[2t_a-2(L+a)}]}=0, \\
        &e^{-(a+b)}\left(1-e^{-2(L-b)}\right)(e^{-(t_a-t_b)} +e^{t_a-t_b})-\frac{2e^{-2a}}{\alpha} -\frac{1}{1+\exp{[(t_a+t_b)-(a+b)}]}+\\
        &\quad+ \frac{1}{1+\exp{[(t_a+t_b)-(2L+a-b)}]}-\frac{1}{1+\exp{[2t_a-2(L+a)}]}=0,
    \end{aligned}
\end{equation}
where we discarded all non-leading terms. We now consider solutions of these equations for three different time regimes: the early times, the intermediate period and the late times.

$\,$

The early time regime is characterized by the hierarchy
\begin{equation}
    \begin{aligned}
        t_a+t_b&\gg a+b,\\
        t_a+t_b&\ll 2L+a-b,\\
        t_a&\ll L+a.
    \end{aligned}
\end{equation}
The first inequality places the direct contribution to the QES equations, while the remaining two keep the time dependent reflected image terms exponentially small. These reflected terms can nevertheless affect the existence and location of the island, because the terms supporting a near-horizon QES are themselves parametrically small. Under these conditions, the QES equations simplify, yielding the following approximate solution for the position and time coordinate of the island endpoint
\begin{equation}\label{eq:IslEarly}
    \begin{aligned}
        t_a&=t_b-\frac{1}{2}\log \left(1-\Sigma_E\right),\\
        a&=\log\frac{1}{\alpha e^{-b}\left(1-e^{-2(L-b)}\right)}-\frac{1}{2}\log \left(1-\Sigma_E\right),
    \end{aligned}
\end{equation}
where 
\begin{equation}
    \Sigma_E=\frac{e^{-2 t_b}}{\alpha^2 e^{-4b} \left(1-e^{-2(L-b)}\right)^3}+\frac{e^{2 t_b}e^{-2(L-b)}}{\left(1-e^{-2(L-b)}\right)}-e^{2t_b}\alpha e^{-2b} e^{-2(L-b)}.
\end{equation}
The late time regime is characterized by
\begin{equation}
    \begin{aligned}
        t_a+t_b&\gg a+b,\\
        t_a+t_b&\gg 2L+a-b,\\
        t_a&\gg L+a.
    \end{aligned}
\end{equation}
In this regime, both reflected-image terms have reached their large time limits. Their leading contributions to the QES equations cancel, leaving exponentially small residual terms. Once these residual terms are sufficiently small, a real near-horizon exterior QES can exist again. The resulting solution has a similar form to the early time solution
\begin{equation}\label{eq:IslLate}
    \begin{aligned}
        t_a&=t_b-\frac{1}{2}\log \left(1-\Sigma_L\right),\\
        a&=\log\frac{1}{\alpha e^{-b}\left(1-e^{-2(L-b)}\right)}-\frac{1}{2}\log \left(1-\Sigma_L\right),
    \end{aligned}
\end{equation}
with 
\begin{equation}
    \Sigma_L=\frac{\left(1-\alpha e^{-2b}\left(1-e^{-2(L-b)}\right)^2\right)e^{-2t_b}}{\alpha^3 e^{-6b} \left(1-e^{-2(L-b)}\right)^4 e^{-2(L-b)}}.
\end{equation}
The early and late time solutions correspond to real quantum extremal surfaces provided that $\Sigma_E<1$ and $\Sigma_L<1$, respectively. To investigate the behavior of the exterior island between these two regimes, we consider the intermediate time hierarchy
\begin{equation}
    \begin{aligned}
        t_a+t_b&\gg a+b,\\
        t_a+t_b&\gg 2L+a-b,\\
        t_a&\ll L+a,
    \end{aligned}
\end{equation}
In this regime, the reflected contribution coupling the island and radiation endpoints is of order unity in the reduced QES equations. The compensating image contribution involving the island endpoints remains exponentially small. As shown in Appendix~\ref{sec:AppBlink}, this imbalance prevents the QES equations from admitting a near-horizon solution in the black hole exterior. The temporary absence of this exterior island branch corresponds to the intermediate stage of the blinking phenomenon.

$\,$

We estimate the disappearance and reappearance times of the exterior island branch based on the conditions for the early and late time solutions, \eqref{eq:IslEarly} and \eqref{eq:IslLate}, respectively, to be real
\begin{equation}
    1-\Sigma_E(t_b)>0,\quad 1-\Sigma_L(t_b)>0.
\end{equation}
The first equation 
\begin{equation}
    1-\Sigma_E(t_{b,\,\text{early}}^\pm)=0
\end{equation}
gives the time of formation $t^-_{b,\,\text{early}}$ and the disappearance $t^+_{b,\,\text{early}}$ of the early island configuration, while the second equation 
\begin{equation}
    1-\Sigma_L(t_{b,\,\text{late}})=0
\end{equation}
gives the time $t_{b,\,\text{late}}$ of appearance of the late island configuration.

For a sufficiently distant wall, the near-horizon exterior island branch is absent in the interval \mbox{$(t^{+}_{b,\,\text{early}},t_{b,\,\text{late}})$}. Neglecting corrections of order $\mathcal{O}(\alpha e^{-2b})$ and $\mathcal{O}(e^{-2(L-b)})$, the endpoints of this interval are
\begin{equation}\label{eq:BlinkTime}
    t^+_{b,\,\text{early}}=(L-b),\quad t_{b,\,\text{late}}=(L-b)+3b+\frac{3}{2}\log\frac{1}{\alpha}.
\end{equation}
The disappearance therefore occurs on the same leading-order time scale as the onset of entropy saturation in the absence of the island. This correspondence directly follows from the no-island entropy. The crossover from linear growth toward saturation occurs when $\cosh 2t_b$ becomes comparable to $\cosh 2(L-b)$, giving $t_b\simeq L-b$.

\begin{figure}[t]
    \centering
    \includegraphics[width=0.49\linewidth]{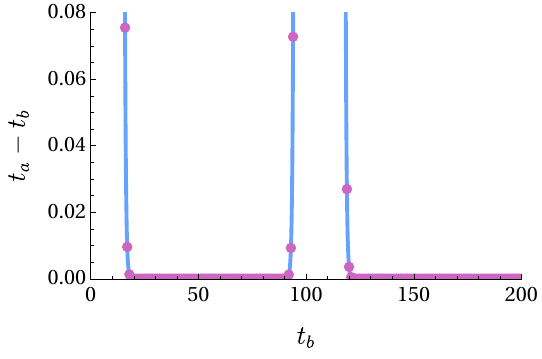}
    \includegraphics[width=0.49\linewidth]{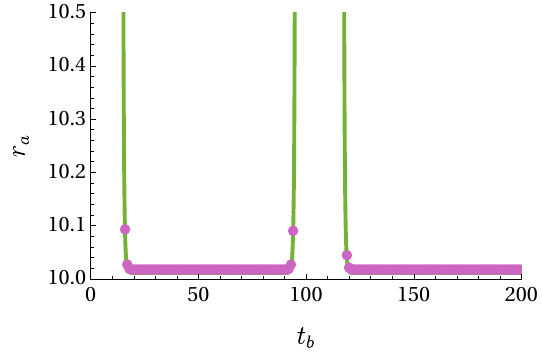}
    \caption{Comparison of the QES solutions found, where the \textit{solid lines} represent the approximate solutions, while the \textit{dots} represent the numerical solutions. The parameters used are $b=5$, $\Phi_h=15$, $L=100$, $c=3$, $G=0.1.$
}
    \label{fig:NumAndApp}
\end{figure}

As shown in Fig.~\ref{fig:NumAndApp}, the approximate solutions agree well with the numerical results for both the position and the time coordinate of the island endpoint within their range of validity.

\begin{figure}[t]
    \centering
    \includegraphics[width=0.45\linewidth]{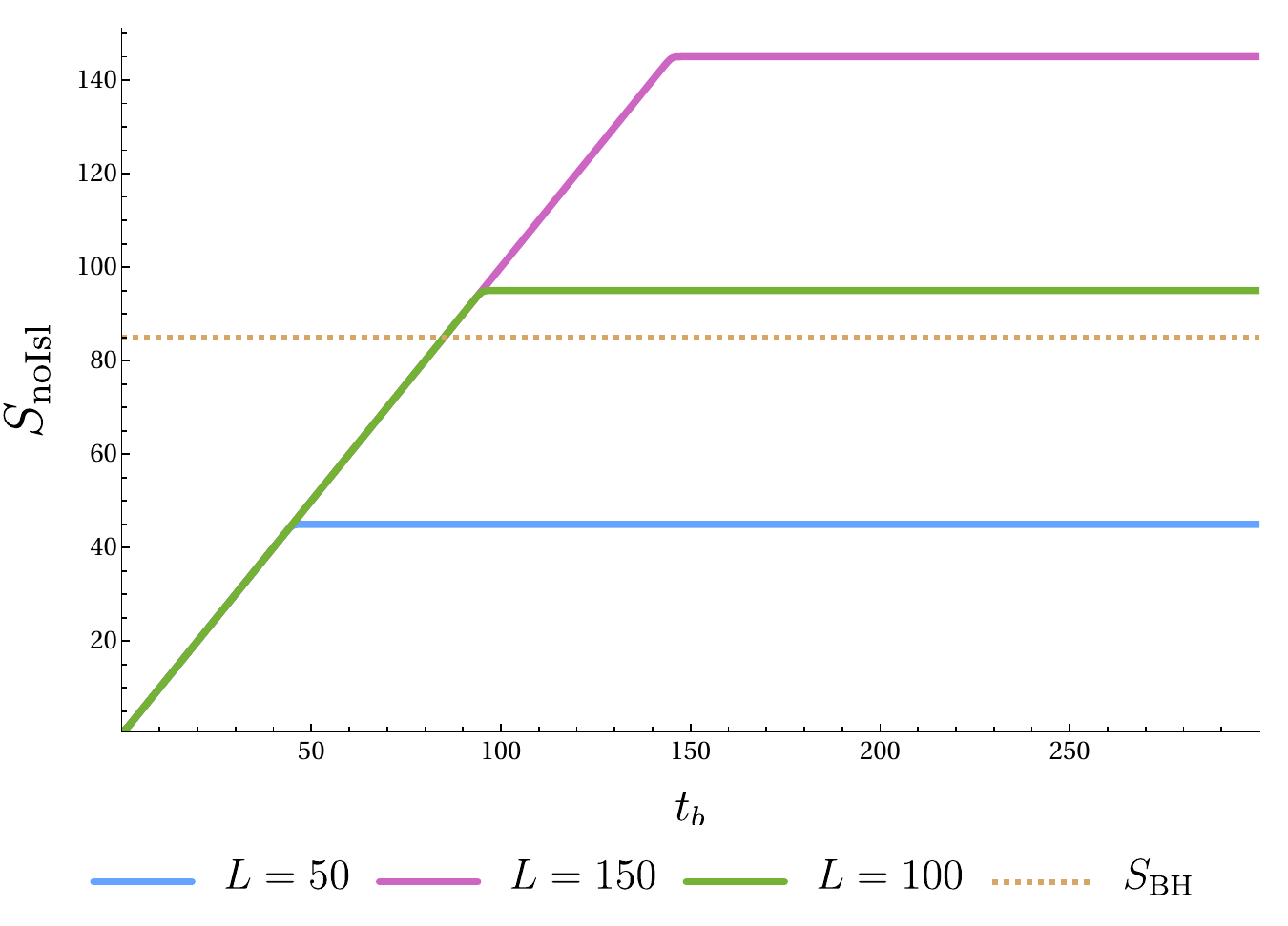}
    \includegraphics[width=0.45\linewidth]{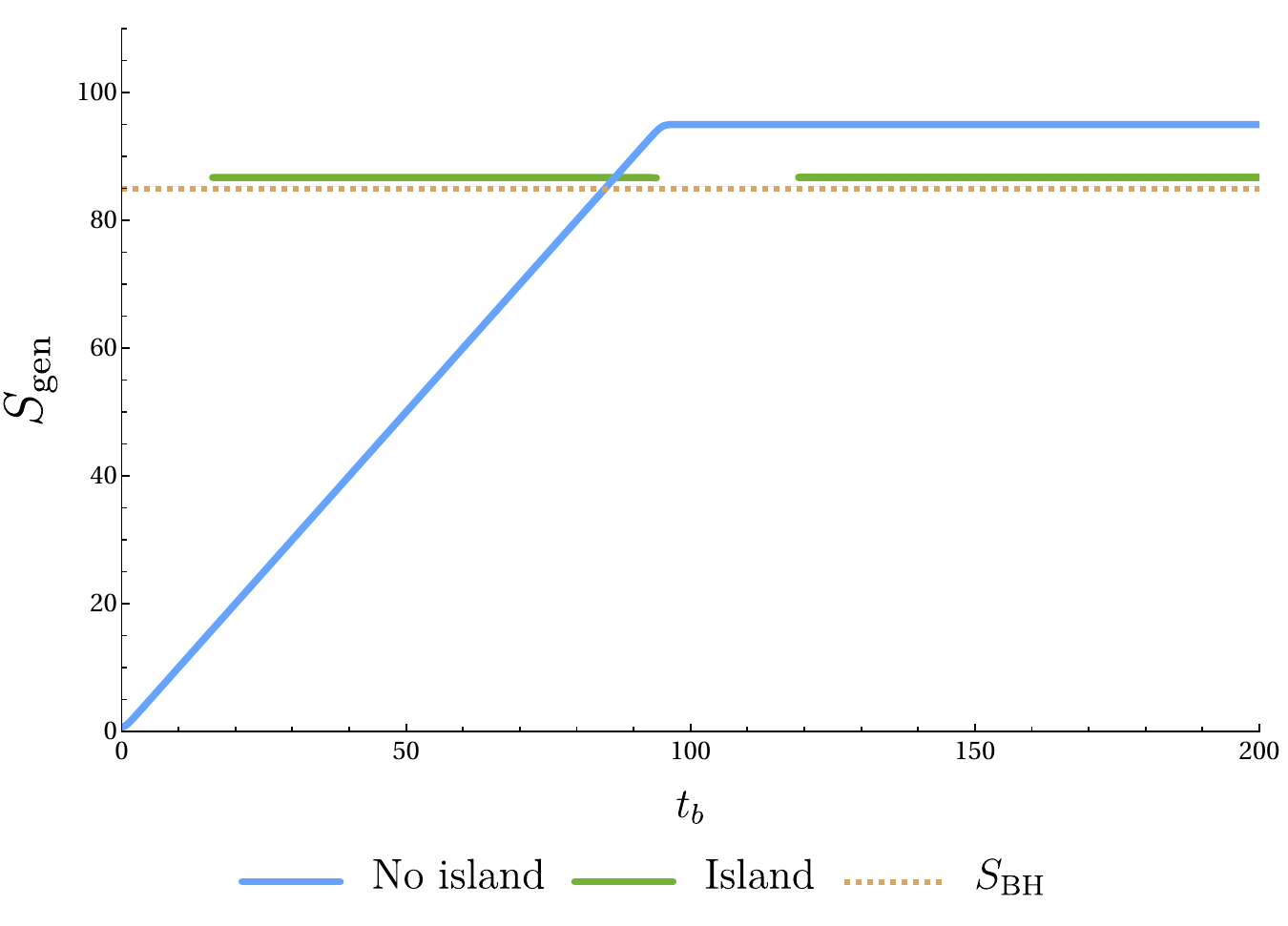}
    \caption{Time evolution of the entanglement entropy without an island for the different positions of the wall (\textit{left}). Time dependence of the entanglement entropy with an island (\textit{right}), where the \textit{red line} denotes the entropy without the island and the \textit{green line} denotes the island configuration. The parameters used are $b=5$, $L=100$, $\Phi_h=15$, $c=3$, $G=0.1$}
    \label{fig:NumericEnt}
\end{figure}

Finally, following the QES prescription, the physical
radiation entropy is obtained by comparing the
no-island contribution with the generalized entropies
of all admissible island extrema, 
\begin{equation}
    S(\mathcal R;t_b)
    =\min\left\{
        S_{\rm noIsl}(t_b),
        S_{\rm isl,min}(t_b)
    \right\},
\end{equation}
where
\begin{equation}
    S_{\rm isl,min}(t_b)
    \equiv
    \min_{\mathcal I\in\mathfrak Q_{\rm adm}(t_b)}
    S_{\rm gen}(\mathcal I;t_b),
\end{equation}
and $\mathfrak Q_{\rm adm}(t_b)$ denotes the set of
admissible quantum extremal surface configurations,
including interior configurations when appropriate.

To understand the role of the finite bath in the
information paradox, it is first useful to
compare the saturated no-island entropy
\eqref{eq:EntSat} with the thermodynamic
black hole entropy, defined as
\begin{equation}\label{eq:thermodynamicEnt}
    S_{\rm BH}
    =\lim_{a\rightarrow\infty}S_{\rm grav}(a)
    =2S_0+\frac{\Phi_h}{2G}.
\end{equation}
In the conventional semiclassical description,
the growth of the radiation entropy beyond the
black hole entropy scale is commonly regarded
as a manifestation of the information paradox.
In our setup, however, the reflecting wall
changes this picture: the no-island entropy
does not grow indefinitely, but instead
approaches the finite value
\begin{equation}
    S_{\rm noIsl}^{\infty}
    =\frac{c}{3}
    \log\left[
        \frac{2\sinh(L-b)}{\kappa\epsilon_b}
    \right].
\end{equation}

If we require that this saturation value does
not exceed the thermodynamic entropy,
\begin{equation}
    S_{\rm noIsl}^{\infty}\leq S_{\rm BH},
\end{equation}
we obtain
\begin{equation}\label{eq:ShortBathThermalBound}
    \sinh(L-b)
    \leq
    \frac{\kappa\epsilon_b}{2}
    \exp\left[
        \frac{3}{c}
        \left(2S_0+\frac{\Phi_h}{2G}\right)
    \right].
\end{equation}
In the regime $L-b\gg1$ this condition becomes
\begin{equation}
    L-b\lesssim
    \frac{3S_{\rm BH}}{c}
    +\log(\kappa\epsilon_b).
\end{equation}
Thus, for a sufficiently short bath, the
no-island entropy saturates below the
thermodynamic black hole entropy, so that
the usual information paradox problem
associated with its growth beyond this scale
does not arise. This is analogous to the
saturation mechanism observed previously
for eternal black holes placed inside finite
cavities~\cite{Ageev:2023hxe}.

However, this comparison with the thermodynamic
entropy is not sufficient to determine which
configuration actually dominates the radiation
entropy. In general, the island prescription
requires comparing the complete generalized
entropies of the competing saddles, rather
than comparing the no-island entropy with
the purely gravitational contribution
$S_{\rm BH}$. In particular, the generalized
entropy of an island contains a nontrivial
matter contribution and does not generally
coincide with the thermodynamic entropy.

For example, in the distant wall regime,
before reflected correlations significantly
modify the exterior island, the generalized
entropy approaches the familiar plateau
\begin{equation}\label{eq:ExteriorPlateauComparison}
    S_{\rm gen}^{\rm ext}
    \simeq
    S_{\rm BH}
    +\frac{c}{3}
    \log\frac{2e^b}{\kappa\epsilon_b},
\end{equation}
where we have suppressed additive boundary and normalization constants. Consequently, the difference between the saturated no-island entropy and this reference island entropy scale is
\begin{equation}
    S_{\rm noIsl}^{\infty}
    -S_{\rm gen}^{\rm ext}
    \simeq
    -S_{\rm BH}
    +\frac{c}{3}
    \log \frac{\sinh(L-b)}{e^b} .
\end{equation}
A formal comparison of these two limiting
expressions, requiring
$S_{\rm noIsl}^{\infty}\lesssim
S_{\rm gen}^{\rm ext}$, gives
\begin{equation}\label{eq:ShortBathIslandBenchmark}
    \sinh(L-b)
    \lesssim
    e^b
    \exp\left(\frac{3S_{\rm BH}}{c}\right).
\end{equation}
For $L-b\gg1$, this can be written as
\begin{equation}
    L-b
    \lesssim
    \frac{3S_{\rm BH}}{c}
    +b+\log 2.
\end{equation}
This illustrates why the thermodynamic
criterion \eqref{eq:ShortBathThermalBound}
and the comparison with a nontrivial
island entropy are different: the latter
contains an additional contribution
from the entanglement of matter fields.

The comparison in
Eq.~\eqref{eq:ShortBathIslandBenchmark}
should nevertheless be understood only
as an estimate of the relevant entropy
scales, not as an exact phase boundary.
Indeed, the no-island saturation value
is reached for $t_b\gg L-b$, whereas
the exterior plateau
\eqref{eq:ExteriorPlateauComparison}
is obtained when the effects of the
reflecting wall are negligible.
The two limiting expressions therefore
do not describe the competing saddles
in a common late time regime.
The actual dominance condition must
be evaluated using the generalized
entropies at the same radiation time.

$\,$

This distinction can also be understood
from the Page transition. Before the
wall contributions become important, in the regime
$1\ll t_b\ll L-b$, the no-island entropy
behaves as
\begin{equation}
    S_{\rm noIsl}(t_b)
    \simeq
    \frac{c}{3}
    \left(t_b-\log\kappa\epsilon_b\right),
\end{equation}
while the exterior-island entropy is
approximately given by
Eq.~\eqref{eq:ExteriorPlateauComparison}.
Equating these two contributions gives the Page time
\begin{equation}\label{eq:PageTimeComparison}
    t_{\rm Page}
    \simeq
    \frac3 c S_{\rm BH}
    +b+\log 2.
\end{equation}
This crossing is physically relevant
only if an admissible island saddle
exists and both entropy approximations
remain valid at the crossing time.

The comparison between the bath scale Page time
therefore provides a useful qualitative
distinction. If \mbox{$L-b\lesssim t_{\rm Page}$}, the reflecting wall modifies the no-island entropy before the ordinary
infinite bath Page transition would
take place. In particular, its
saturation may prevent the usual
unbounded entropy growth without
requiring an island to become dominant
at that time. Conversely, if \mbox{$L-b\gtrsim t_{\rm Page}$},
a Page transition can occur before
the reflected correlations become
important, provided the corresponding
QES exists.

These observations emphasize that
the saturation of the no-island entropy,
the existence of a quantum extremal
surface, and the dominance of the
island contribution are distinct
properties. In particular, the
thermodynamic inequality
\eqref{eq:ShortBathThermalBound}
is a criterion for the absence of
an excess of the no-island entropy
over $S_{\rm BH}$, but is not by
itself a criterion for no-island
dominance.

For a sufficiently distant wall,
the no-island entropy can exceed
the thermodynamic black hole entropy,
giving an information paradox
problem if the trivial configuration
is considered in isolation. 
During the blinking interval
\eqref{eq:BlinkTime}, this issue
becomes especially apparent because
the near-horizon exterior QES branch
temporarily disappears. An analysis
restricted to exterior configurations
may then incorrectly suggest that
the no-island saddle is the only
available contribution.

However, the absence of an exterior
QES does not imply the absence of
other island extrema. As we show
in Section~\ref{iqes:section},
the generalized entropy admits
real future-interior extrema.
For the parameter regime studied
numerically, the interior island
contribution remains below the
no-island entropy throughout the
exterior blinking interval.
Thus, including interior QES
candidates removes the
reappearance of the information
paradox within the class of
configurations investigated here.

In the infinite bath limit \mbox{$(L-b)\rightarrow\infty$} the time of the island disappearance and also the late return time becomes infinite, i.e.
\begin{equation}
    t^+_{b,\,\text{early}}, t_{b,\,\text{late}} \rightarrow \infty,
\end{equation}
and the early island solution is given by  
\begin{equation}
    a\approx b + \log\frac{1}{\alpha}-\frac{1}{2}\log\left(1-\frac{1}{\alpha^2}e^{-2 t_b} e^{4 b}\right)
\end{equation}
with the formation time
\begin{equation}\label{eq:formationTime}
    t^-_{b,\,\text{early}}\approx\log\frac{1}{\alpha}+2 b.
\end{equation}
It is interesting to notice that, in the limit of an infinite bath, the blinking interval does not collapse, but the time of island disappearance is translated to arbitrarily late times. Therefore, the conventional island picture is restored at every fixed $t_b$.


\subsection{Qualitative evolution of QES and scrambling-like behavior}
\label{sec:UnifiedBlinking}
The early and late time exterior solutions derived above have useful unified interpretations. In particular, the difference  between the two branches is not fundamental at the level of the near-horizon equations. Instead, they both arise from a single time-dependent QES existence condition whose different asymptotic limits describe the formation, disappearance, and eventual reappearance of the island. 

Before going into the detailed analysis let us outline the main points to be discussed. We find that the ordinary near-horizon QES first appears at a scrambling-like scale \mbox{$t_{\rm QES}\simeq 2b +\log1/\alpha$}, while the effect of the reflective wall becomes significant at a time \mbox{\(t_{\rm dis}\simeq L-b\)}, proportional to the distance from it. The latter time also controls the onset of saturation of the no-island entropy, whose late time value is set by the thermal entropy of the two finite bath segments, up to ultraviolet and boundary terms. Once the reflected contribution becomes order one in the QES equations, the real symmetric near-horizon exterior solution disappears. It reappears only after a parametrically large delay, giving \mbox{$ \Delta t_{\rm blink}\simeq 3b+3/2\log 1/\alpha$}. So the wall position determines when the blinking interval begins, and its leading duration is independent of \(L\) and contains a scrambling-like logarithm. This analysis also allows us to calculate the critical size of the bath, below which the early branch of the island solution does not occur at all.

These observations clarify two useful limits. First, the QES formation time and the Page time are parametrically distinct. The formation time is \mbox{$t_{\rm QES}\sim\log 1/\alpha$}, while the Page time is typically of order \mbox{$t_{\rm Page}\sim 1/\alpha$}. Therefore, a QES may blink without ever becoming the dominant entropy saddle. Second, the infinite bath limit is not uniform. Indeed, as \mbox{$L\to\infty$}, the blinking interval does not collapse, but is shifted to arbitrarily later times, and the conventional infinite bath solution is recovered for every fixed time of the radiation region.

Finally, the wall dependence is introduced through the non-local BCFT image contributions, while the local Hartle-Hawking stress tensor in the gravitating region has no corresponding $L$-dependence, emphasizing that the transition is primarily driven by a reorganization of entanglement, rather than a local wall-induced backreaction.

$\,$

It is convenient to introduce some new notation for the variables and parameters used in QES equations \eqref{eq:QESeqFull},
\begin{equation}
    \begin{aligned}
        T&=e^{t_a-t_b}\\
        d&=L-b\\
        F&=1-e^{-2d}
    \end{aligned},\quad 
    \begin{aligned}
        X&=e^{t_a+t_b-(a+b)}\\
        Y&=e^{t_a+t_b-(2L+a-b)}\\
        Z&=e^{2t_a-2(L+a)}
    \end{aligned}\, .
\end{equation}
A particularly simple relation follows by adding the two reduced QES
equations.  All terms containing $X$, $Y$, and $Z$ cancel pairwise, giving
\begin{equation}
\label{eq:UnifiedRadialRelation}
    e^{-a} = \frac{\alpha e^{-b}F}{T}.
\end{equation}
Thus, throughout the near-horizon regime,
\begin{equation}
\label{eq:UnifiedAT}
    a=a_0+\log T,
    \qquad
    t_a=t_b+\log T,
    \qquad
    a_0\equiv
    \log\frac{1}{\alpha e^{-b}F}.
\end{equation}
The overall structure of the early and late time exterior solutions directly follows from the reduced QES equations and is not a result of specific asymptotic approximations.

By substituting \eqref{eq:UnifiedRadialRelation} into the expressions for $X, Y$ and $Z$, we can obtain them as functions that depend only on the time $t_b$ of the radiation region,
\begin{equation}
\label{eq:XYZTimeOnly}
    X=\alpha F e^{2t_b-2b},\qquad
    Y=e^{-2d}X,\qquad
    Z=\alpha F e^{-2L}X.
\end{equation}
Then the remaining QES equation can be written in the compact form
\begin{equation}
\label{eq:MasterQES}
    T^{-2} =1-\frac{\mathcal H(t_b)}{p},\quad 
    p\equiv \alpha e^{-2b}F^2,
\end{equation}
where
\begin{equation}
\label{eq:HDef}
    \mathcal H(t_b)\equiv\frac{1}{1+X}
    +\frac{Y}{1+Y} -\frac{Z}{1+Z}.
\end{equation}
Consequently, a real symmetric near-horizon QES exists when
\begin{equation}
\label{eq:QESExistenceCriterion}
    \mathcal H(t_b)<p.
\end{equation}
The corresponding spacetime location is
\begin{equation}
\label{eq:UnifiedQESSolution}
    t_a = t_b -\frac12\log\left(1-\frac{\mathcal H(t_b)}{p}
    \right), \qquad
    a = a_0 -\frac12\log\left(1-\frac{\mathcal H(t_b)}{p}
    \right).
\end{equation}
The derivation and the relation to $\Sigma_E$ and $\Sigma_L$ from the previous section are given in Appendix~\ref{sec:AppUnifiedBlinking}.

Equation~\eqref{eq:QESExistenceCriterion} gives a simple physical explanation of the blinking phenomenon. The first term in the $\mathcal{H}$ is the residual contribution from the ordinary, wall-independent setup. It decreases exponentially as the near-horizon QES develops. The second term is generated by the reflected image and increases as the wall enters the relevant causal region. The third term is another image contribution involving the island endpoint, and it enters with the opposite sign. Thus, the island undergoes three stages: the initial dominance of wall-independent contribution, followed by the development of wall contribution, and finally, the mutual reduction of the image contributions.

In the early time regime one has \mbox{$\mathcal H/p \simeq\Sigma_E$}, whereas at late times \mbox{$\mathcal H/p \simeq\Sigma_L$}. In the intermediate regime,
\begin{equation}
    X\gg1, \quad  Y\gg1, \quad Z\ll1,
\end{equation}
and therefore \mbox{$\mathcal H\simeq1$}. Since $p=\alpha e^{-2b}F^2\ll1$ in the semiclassical regime, condition~\eqref{eq:QESExistenceCriterion} cannot be satisfied. The absence of the intermediate near-horizon exterior solution can thus be understood directly as a mismatch between an order-one reflected contribution and the parametrically small gravitational scale $p$.

$\,$

The same formulation allows us to estimate the minimum bath size required for an early time island branch to exist. For a sufficiently small bath, only the late time branch is present. An early solution branch exists only if condition \eqref{eq:QESExistenceCriterion} has two distinct roots. As shown in Appendix~\ref{sec:AppUnifiedBlinking}, this requires
\begin{equation}
\label{eq:CriticalBathExact}
    4\,\frac{e^{4b-2d}}{\alpha^2F^3}\left(
        \frac{1}{F}-\alpha e^{-2b}\right)<1.
\end{equation}
For $d=L-b\gg1$, so that $F\simeq1$, this reduces to the simple
criterion
\begin{equation}
\label{eq:CriticalBathApprox}
    d>d_c,\quad d_c\simeq 2b+\log\frac{2}{\alpha}.
\end{equation}
Close to the critical value, the time duration during which the early branch exists can be estimated as follows 
\begin{equation}
    \Delta t_{\rm early} \sim \sqrt{d-d_c}.
\end{equation}

Another useful consequence of Eq.~\eqref{eq:BlinkTime} is that, to leading order, the duration of the blinking interval is
\begin{equation}
\label{eq:BlinkDuration}
    \Delta t_{\rm blink} \equiv t_{b,\mathrm{late}}
    - t^+_{b,\mathrm{early}}\simeq
    3b+\frac32\log\frac{1}{\alpha}.
\end{equation}
Remarkably, the leading dependence on the wall position cancels.  The bath size determines when the blinking phenomenon is occurred,
whereas its duration is controlled by the radiation endpoint and the
semiclassical parameter.  In physical units, this separation of scales translates to
\begin{equation}
    t_{\rm dis}^{\rm phys}\equiv t_{b,\, \rm early}^{+\,\,\rm phys}\simeq \frac{\beta}{2\pi}\left(L-b\right),
    \qquad \Delta t_{\,\rm blink}^{\,\rm phys} \simeq
    \frac{3\beta}{2\pi}b + \frac{3\beta}{4\pi} \log\frac{1}{\alpha}.
\end{equation}
Giving that
\begin{equation}
    \frac{1}{\alpha}=\frac{3 \Phi_h}{c G}=\frac{6}{c}\left(S_{\mathrm{BH}}-2 S_0\right),
\end{equation}
we can rewrite the physical blinking duration in explicit scrambling-like form
\begin{equation}
    \Delta t_{\mathrm{blink}}^{\mathrm{phys}} \simeq \frac{3\beta}{2\pi}b+\frac{3 \beta}{4 \pi} \log \left[\frac{6}{c}\left(S_{\mathrm{BH}}-2 S_0\right)\right] .
\end{equation}
The onset of blinking is therefore controlled by the distance between the reflecting wall and the radiation region, as well as by the point where the contribution from the wall becomes the leading order in the QES equations. After the reflected contribution becomes dominant, the system requires an additional, parameterically long time period for the competing image contributions to reorganize so that a real near-horizon extremum can form again, which leads to the subsequent reappearance of the island. This additional delay is logarithmic in the inverse semiclassical parameter and is therefore naturally of the scrambling type~\cite{Hayden:2007cs,Sekino:2008he,Batista:2026slh}. In this sense, two distinct scales related to scrambling appear in the problem. The ordinary island itself develops only after a familiar logarithmic delay, which is associated with the emergence of a near-horizon QES. The late island is then restored only after a second logarithmic delay that follows the wall-induced disappearance of the early branch.

This separation of scales provides a useful physical picture of the blinking phenomenon. The reflecting wall determines when the system is driven away from the standard infinite bath island configuration, through a causal propagation time fixed by the size of the bath. The recovery of the island is then governed by the semiclassical information-processing scale of the black hole. Therefore, increasing the wall distance primarily translates the blinking episode to later times without changing its leading duration, while taking the semiclassical limit makes the recovery parametrically slow. The blink may consequently be viewed as an interplay between an infrared propagation scale associated with the finite bath and a scrambling-like scale associated with the near-horizon quantum extremal surface. This also explains why the infinite bath limit is recovered by pushing the entire phenomenon to arbitrarily late times rather than by shrinking the blinking interval itself.

$\,$

The existence of an exterior QES does not necessarily imply its dominance in the radiation entropy. Indeed, the formation time of the near-horizon QES, given by Eq.~\eqref{eq:formationTime}, is parametrically earlier than the Page time estimated in Eq.~\eqref{eq:PageTimeComparison}. For fixed $b$ and in the semiclassical regime, these scales behave as 
\begin{equation}
    t_{\rm QES}\sim\log \frac 1\alpha,\qquad t_{\rm Page}\sim \frac 1\alpha,
\end{equation}
respectively, provided that $S_0$ does not introduce a parametrically larger scale. Thus, an exterior QES may exist and even undergo blinking without ever becoming the dominant entropy saddle.

This distinction is particularly important in the presence of reflecting boundaries. If the no-island entropy saturates before the Page transition, the disappearance of an exterior QES need not affect the radiation entropy. If, however, an island has already become dominant, the loss of the exterior branch raises a different question: whether the entropy is then governed by the no-island configuration or by QESs outside the exterior region. This motivates extending the extremization into the black hole interior, as we do in the next section.

These observations make it clear that the wall affects the island primarily through the nonlocal entanglement structure of the BCFT. In particular, as discussed in Sec.~\ref{sec:Conclusion}, the local Hartle--Hawking stress tensor in the gravitating region carries no explicit dependence on $L$, whereas the QES existence condition depends sensitively on the wall through the image terms. The blinking transition is therefore caused by the reorganization of long range correlations, which contribute to the overall entropy.

\section{Quantum extremal surfaces in the black hole interior}
\label{iqes:section}

The absence of a real exterior solution during the blinking interval does not by itself imply the absence of a quantum extremal surface. For example dynamical horizon crossing of a QES in evaporating JT gravity~\cite{Hollowood:2020cou} suggests that the horizon  does not necessarily terminate the QES branch, promting a search for its possible continuation into the interior Kruskal region. The multiple competing island saddles found in~\cite{Hollowood:2020kvk} (see also \cite{Hollowood:2021wkw, Hollowood:2021nlo,Gyongyosi:2022vaf,Craps:2022ahp,Batista:2026slh}) further illustrate why the behavior of a single exterior branch is not enough to determine the radiation entropy. Other extrema may contribute and must be compared through their generalized entropies. Finally, the Schwinger–Keldysh replica construction of Lorentzian islands in~\cite{Goto:2020wnk} in evaporating black hole provides  basis for considering interior QES within the island prescription in the regions where purely Euclidean description may fail. Although these studies focus on evaporating black holes, they motivate us to investigate whether the blinking in our finite bath setup reflects a change in the position of island under the horizon or dominance of the other QES branches.

In this section we construct the generalized entropy for symmetric interior configurations, retaining the reflecting wall contributions, and derive the corresponding QES equations. We then obtain a controlled analytical approximation and construct numerical solutions, using regular Kruskal coordinates near horizon crossing. Finally, we compare the interior, exterior and no-island entropy branches to determine whether interior extrema maintain an island contribution throughout the exterior blinking interval.

$\,$

The exterior coordinates used in Section~\ref{sec:Islands} cover only part of the spacetime. Indeed, Eqs.~\eqref{eq:UnifiedAT} and~\eqref{eq:MasterQES} give the right island endpoint as
\begin{equation}
  w_1=e^{t_b-a_0},
  \qquad
  \bar w_1=e^{-a_0-t_b}\left(1-\frac{\mathcal{H}(t_b)}{p}\right),
  \label{eq:int-horizon}
\end{equation}
where $\mathcal H(t_b)$ is defined in Eq.~\eqref{eq:HDef}. As $\mathcal H(t_b)$ approaches $p$, $a$ and $t_a$ diverge, but $w_1$ remains finite and $\bar w_1$ tends to zero. The endpoint therefore approaches the future horizon. The sign change of $\bar w_1$ for $\mathcal H(t_b)>p$ suggests that the extremum can continue into the future interior. To examine this continuation, we solve the same QES stationarity conditions as in the exterior calculation,
\begin{equation}
  \frac{\partial S_{\rm gen}}{\partial a}=0,
  \qquad
  \frac{\partial S_{\rm gen}}{\partial t_a}=0,
  \label{eq:int-qes-conditions}
\end{equation}
with the radiation endpoints held fixed as in Eq.~\eqref{eq:RadiationEndpoints}. We now continue both island endpoints, preserving their left-right symmetry, and construct the generalized entropy in the interior region.
We keep the dimensionless thermal coordinates, the semiclassical regime \mbox{$\alpha\ll1$}, and the renormalization prescription of the exterior calculation. Common additive boundary terms are suppressed. For the future-interior configuration island coordinates are written explicitly
\begin{equation}
  \begin{aligned}
    P_1&=(e^{-A+\tau},-e^{-A-\tau}),\\
    P_3&=(e^{-A-\tau},-e^{-A+\tau}),
    \qquad A>0,\quad \tau>0 .
  \end{aligned}
  \label{eq:int-endpoints}
\end{equation}
Thus $w_1\bar w_1=-q^2$, where $q=e^{-A}$, and both endpoints lie in the future-interior quadrant. The dilaton~\eqref{eq:Dilaton} becomes
\begin{equation}
  \Phi(P_1)=\Phi(P_3)
  =\Phi_h\frac{1-q^2}{1+q^2}
  =\Phi_h\tanh A .
  \label{eq:int-dilaton}
\end{equation}
The same change of patch follows from the formal continuation
\begin{equation}
  a=A+\frac{i\pi}{2},
  \qquad
  t_a=\tau+\frac{i\pi}{2}.
  \label{eq:int-continuation}
\end{equation}
Applied simultaneously to both island endpoints, this continuation gives the real coordinates in Eq.~\eqref{eq:int-endpoints}. An analytically continued QES solution branch satisfies the extremality conditions of the continued generalized entropy. In the interior variables, Eq.~\eqref{eq:int-qes-conditions} therefore can be rewritten as 
\begin{equation}
    \frac{\partial S_{\rm gen}^{F}}{\partial A}=0,\qquad 
    \frac{\partial S_{\rm gen}^{F}}{\partial \tau}=0,
\end{equation}
where $S_{\rm gen}^{F}$ denotes the continued generalized entropy. A continued solution with real $A$ and $\tau$ consequently describes a QES in the future interior, even though the exterior coordinates $a$ and $t_a$ are complex. The coordinate $\tau$ is a spatial boost coordinate in this region, rather than a time coordinate.

$\,$

Evaluating the free-Dirac BCFT entropy at these endpoints, or equivalently
continuing Eqs.~\eqref{eq:exteriorIslandCFT0}--\eqref{eq:EntGen}, gives
\begin{equation}
  \begin{aligned}
    S_{\mathrm{gen}}^F(A,\tau;t_b)
    ={}&2S_0+\frac{\Phi_h}{2G}\tanh A\\
    &+\frac{c}{3}\log\!\left[
      \frac{4\sinh\tau\cosh t_b}{\kappa\epsilon_b\cosh A}
      \frac{\sinh(A+b)-\sinh(\tau-t_b)}
           {\sinh(A+b)+\sinh(\tau+t_b)}\right]\\
    &+\frac{c}{3}\log\!\left[
      \frac{\sinh(2L+A-b)+\sinh(\tau+t_b)}
           {\sinh(2L+A-b)-\sinh(\tau-t_b)}\right]\\
    &+\frac{c}{6}\log\!\left[
      \frac{4\cosh^2(L+A)\sinh^2 d}
           {[\cosh 2(L+A)+\cosh 2\tau]
            [\cosh 2d+\cosh 2t_b]}\right],
  \end{aligned}
  \label{eq:int-entropy}
\end{equation}
where $d=L-b$, and the superscript $F$ denotes the future
interior. The logarithms are evaluated on the real branch corresponding
to the spacelike interval configuration. The last two terms retain the
wall-image contributions.

At fixed $t_b$, the QES conditions are obtained by differentiating
Eq.~\eqref{eq:int-entropy}. After division by $c/3$, they read
\begin{equation}
  \begin{aligned}
    0={}&\coth\tau
      -\frac{\cosh(\tau-t_b)}{\sinh(A+b)-\sinh(\tau-t_b)}
      -\frac{\cosh(\tau+t_b)}{\sinh(A+b)+\sinh(\tau+t_b)}\\
    &+\frac{\cosh(\tau+t_b)}{\sinh(2L+A-b)+\sinh(\tau+t_b)}
      +\frac{\cosh(\tau-t_b)}{\sinh(2L+A-b)-\sinh(\tau-t_b)}\\
    &-\frac{\sinh 2\tau}{\cosh 2(L+A)+\cosh 2\tau},
  \end{aligned}
  \label{eq:int-qes-tau}
\end{equation}
and
\begin{equation}
  \begin{aligned}
    0={}&\frac{1}{2\alpha\cosh^2 A}
      +\tanh(L+A)-\tanh A
      -\frac{\sinh 2(L+A)}{\cosh 2(L+A)+\cosh 2\tau}\\
    &+\frac{\cosh(A+b)}{\sinh(A+b)-\sinh(\tau-t_b)}
      -\frac{\cosh(A+b)}{\sinh(A+b)+\sinh(\tau+t_b)}\\
    &+\frac{\cosh(2L+A-b)}{\sinh(2L+A-b)+\sinh(\tau+t_b)}
      -\frac{\cosh(2L+A-b)}{\sinh(2L+A-b)-\sinh(\tau-t_b)}.
  \end{aligned}
  \label{eq:int-qes-A}
\end{equation}
In particular, the dilaton contribution to the radial derivative is now
$\Phi_h/(2G\cosh^2 A)$, with the opposite sign to its exterior counterpart
in Eq.~\eqref{eq:QESRadial}.

For $q>0$, the Jacobian
\begin{equation}
  \det\frac{\partial(w_1,\bar w_1)}{\partial(A,\tau)}
  =2w_1\bar w_1=-2q^2
  \label{eq:int-jacobian}
\end{equation}
does not vanish. Equations~\eqref{eq:int-qes-tau} and~\eqref{eq:int-qes-A}
are therefore equivalent to extremizing with respect to the two Kruskal
coordinates in the symmetric sector. At the horizon, however, this chart
degenerates. Numerical continuation across the horizon should instead use
the Kruskal form of the same functional:
\begin{equation}
  \begin{aligned}
    S_{\mathrm{gen}}^{\mathrm{sym}}
    ={}&2S_0+\frac{\Phi_h}{2G}
               \frac{1+w_1\bar w_1}{1-w_1\bar w_1}\\
    &+\frac{c}{3}\log\!\left[
      \frac{2(w_2-w_1)(\bar w_2-\bar w_1)
               (w_2+\bar w_2)(w_1+\bar w_1)}
           {\kappa\epsilon_b e^b(1-w_1\bar w_1)
               (w_1+\bar w_2)(\bar w_1+w_2)}\right]\\
    &+\frac{c}{3}\log\!\left[
      \frac{(R^2-w_1\bar w_1)(R^2-e^{2b})
               (R^2+w_1w_2)(R^2+\bar w_1\bar w_2)}
           {(R^2-w_1\bar w_2)(R^2-\bar w_1w_2)
            \sqrt{(R^2+w_1^2)(R^2+\bar w_1^2)
                   (R^2+w_2^2)(R^2+\bar w_2^2)}}\right].
  \end{aligned}
  \label{eq:int-kruskal-entropy}
\end{equation}
Here $R=e^L$ is the wall radius introduced in Section~2. This expression
is regular at $\bar w_1=0$ provided no direct or image separation vanishes.
Crossing the horizon is therefore not itself a singularity of the entropy.

\subsection{Controlled interior solution}

We next derive an approximate solution after the reflected contribution
becomes significant. In addition to $\alpha\ll1$, assume
\begin{equation}
  \begin{gathered}
    b,\ d,\ A,\ \tau,\ t_b\gg1,\qquad t_b-\tau\gg1,\\
    \tau+t_b\gg A+b,\qquad L+A-\tau\gg1.
  \end{gathered}
  \label{eq:int-hierarchy}
\end{equation}
On the solution below, $t_b-\tau\gg1$ becomes $t_b-d\gg1$, so this approximation excludes the narrow horizon-crossing region. We keep the combination $q\,e^{t_b-L}$ unexpanded, as well as the exact dilaton and the $q$-dependent Weyl factor.

Using $S_{\mathrm{BH}}$ defined in Eq.~\eqref{eq:thermodynamicEnt}, the leading entropy is therefore
\begin{equation}
  \begin{aligned}
    S_{\mathrm{app}}=S_{\mathrm{BH}}+\frac{c}{3}\Bigg[
      &b+\log\frac{2}{\kappa\epsilon_b}-(t_b-d)
       -\log(1+q^2)-\frac{q^2}{\alpha(1+q^2)}\\
      &+\log\!\left(1+q e^{t_b-L+(\tau-d)}\right)
       +\log\!\left(1+q e^{t_b-L-(\tau-d)}\right)
    \Bigg].
  \end{aligned}
  \label{eq:int-leading-entropy}
\end{equation}
The term proportional to $1/\alpha$ is the exact difference between the interior dilaton entropy and its horizon value. The spatial equation is
\begin{equation}
  0\simeq\frac{3}{c}\,\partial_\tau S_{\mathrm{app}}
   =\frac{q e^{t_b-L+(\tau-d)}}{1+q e^{t_b-L+(\tau-d)}}
    -\frac{q e^{t_b-L-(\tau-d)}}{1+q e^{t_b-L-(\tau-d)}}.
  \label{eq:int-spatial-approx}
\end{equation}
Since $z/(1+z)$ is strictly increasing for $z>0$, this fixes
\begin{equation}
  \tau\simeq d=L-b.
  \label{eq:int-spatial-solution}
\end{equation}
The remaining equation becomes
\begin{equation}
  \begin{aligned}
    0\simeq{}&\frac{1}{2\alpha\cosh^2 A}
      -\tanh A+\frac{1-q e^{t_b-L}}{1+q e^{t_b-L}}\\
    ={}&\frac{2q^2}{\alpha(1+q^2)^2}
      -\frac{1-q^2}{1+q^2}
      +\frac{1-q e^{t_b-L}}{1+q e^{t_b-L}}.
  \end{aligned}
  \label{eq:int-radial-approx}
\end{equation}
Multiplying by the positive denominator and dividing by $2q$ gives
\begin{equation}
  \alpha q^3+e^{t_b-L}(1-\alpha)q^2
       +(1+\alpha)q-\alpha e^{t_b-L}=0.
  \label{eq:int-cubic}
\end{equation}
For $0<\alpha<1$, the left hand side is negative at $q=0$, tends to
$+\infty$ as $q\to+\infty$, and has derivative
\begin{equation}
  3\alpha q^2+2e^{t_b-L}(1-\alpha)q+1+\alpha>0
  \qquad(q>0).
  \label{eq:int-cubic-monotonicity}
\end{equation}
The reduced equation thus has exactly one positive root. At
$q^2=\alpha/(1-\alpha)$, the terms proportional to $e^{t_b-L}$ cancel
and the remaining terms are positive, so the root satisfies
\begin{equation}
  0<q^2<\frac{\alpha}{1-\alpha}.
  \label{eq:int-root-bound}
\end{equation}
The reduced equation has two useful formal limits. At fixed $\alpha$ and
$e^{t_b-L}\ll1$,
\begin{equation}
  q=\frac{\alpha}{1+\alpha}e^{t_b-L}
       +O\!\left(e^{3(t_b-L)}\right).
  \label{eq:int-small-limit}
\end{equation}
As $e^{t_b-L}\to\infty$,
\begin{equation}
  q^2\longrightarrow\frac{\alpha}{1-\alpha},
  \qquad
  A\longrightarrow\frac12\log\frac{1-\alpha}{\alpha}.
  \label{eq:int-large-limit}
\end{equation}
For small $\alpha$, the latter behavior requires
$e^{t_b-L}\sqrt{\alpha}\gg1$. 

The on-shell entropy follows by setting $\tau=d$ in
Eq.~\eqref{eq:int-leading-entropy} and choosing the positive root of
Eq.~\eqref{eq:int-cubic}. Since the endpoint is extremal, its implicit
variation does not contribute to the time derivative:
\begin{equation}
  \frac{dS_F}{dt_b}
  \simeq\left.\frac{\partial S_{\mathrm{app}}}{\partial t_b}\right|_{A,\tau}
  =\frac{c}{3}\frac{q e^{t_b-L}-1}{q e^{t_b-L}+1}.
  \label{eq:int-entropy-slope}
\end{equation}
Here $S_F$ denotes the entropy evaluated on this interior branch. It
decreases until $q e^{t_b-L}=1$ and then increases. Combining this condition
with Eq.~\eqref{eq:int-cubic} gives the minimum at
\begin{equation}
  \begin{aligned}
    q_*^2&=\frac{\sqrt{1+\alpha^2}-1}{\alpha}
          =\frac{\alpha}{\sqrt{1+\alpha^2}+1},\\
    t_*&=L-\log q_*
         =L+\frac12\log\frac{2}{\alpha}+O(\alpha^2).
  \end{aligned}
  \label{eq:int-entropy-minimum-location}
\end{equation}
The corresponding entropy is
\begin{equation}
  S_{F,\min}\simeq S_{\mathrm{BH}}+\frac{c}{3}\left[
    \log\frac{8q_*}{\kappa\epsilon_b}
    -\log(1+q_*^2)
    -\frac{q_*^2}{\alpha(1+q_*^2)}\right].
  \label{eq:int-entropy-minimum}
\end{equation}
Combining Eq.~\eqref{eq:int-large-limit} with the on-shell form of
Eq.~\eqref{eq:int-leading-entropy} gives the formal late time behavior
\begin{equation}
  S_F\simeq S_{\mathrm{BH}}+\frac{c}{3}\left[
    t_b-L+\log\frac{2\alpha}{\kappa\epsilon_b}-1\right].
  \label{eq:int-entropy-late}
\end{equation}
Finally, the constant backreaction correction found in Section~\ref{sec:backreaction} adds $c/6$ to every two-endpoint island entropy and leaves the QES equations unchanged. 

\subsection{Numerical construction of interior quantum extremal surfaces}

For the numerical calculation, we solve Eqs.~\eqref{eq:int-qes-tau} and~\eqref{eq:int-qes-A} at fixed radiation time $t_b$ and evaluate each admissible extremum using the exact entropy in Eq.~\eqref{eq:int-entropy}. The approximate solution $A=-\log q$, $\tau=d$, with $q$ determined by Eq.~\eqref{eq:int-cubic}, provides a starting approximation for the near-horizon interior branch. Close to horizon crossing, we instead use Eq.~\eqref{eq:int-kruskal-entropy}, where the endpoint coordinates remain regular.

\begin{figure}[t]
    \centering
    \includegraphics[width=0.48\linewidth]{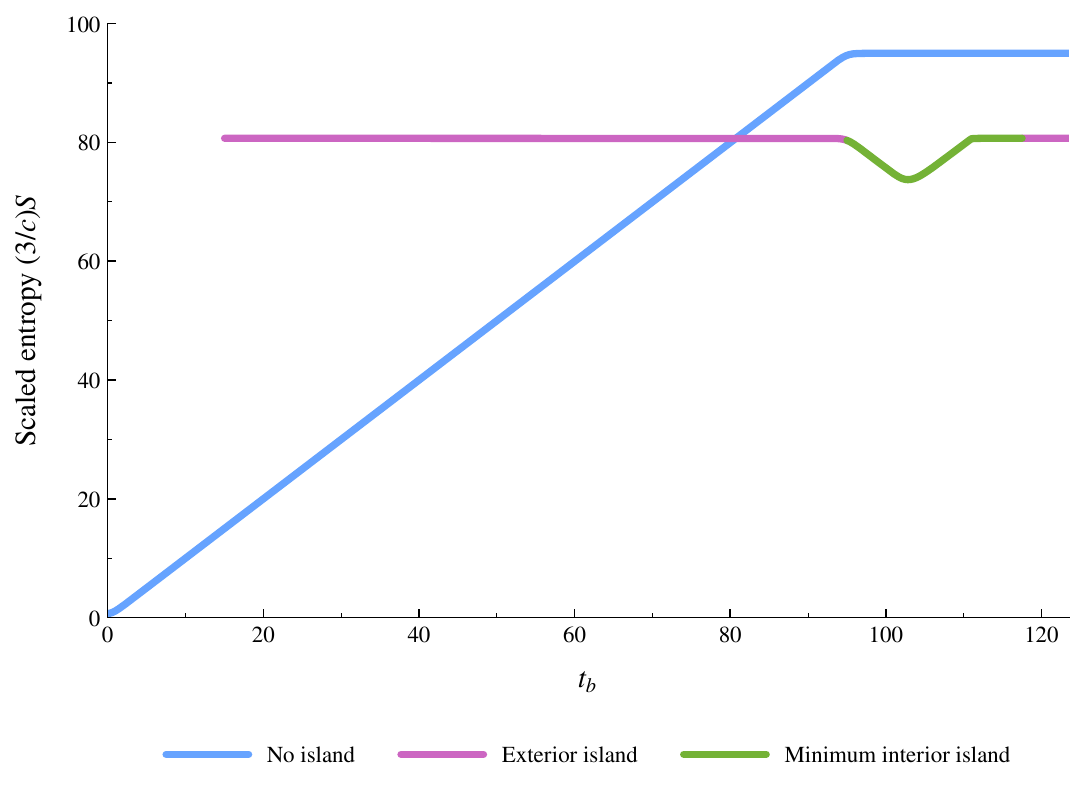}
    \includegraphics[width=0.48\linewidth]{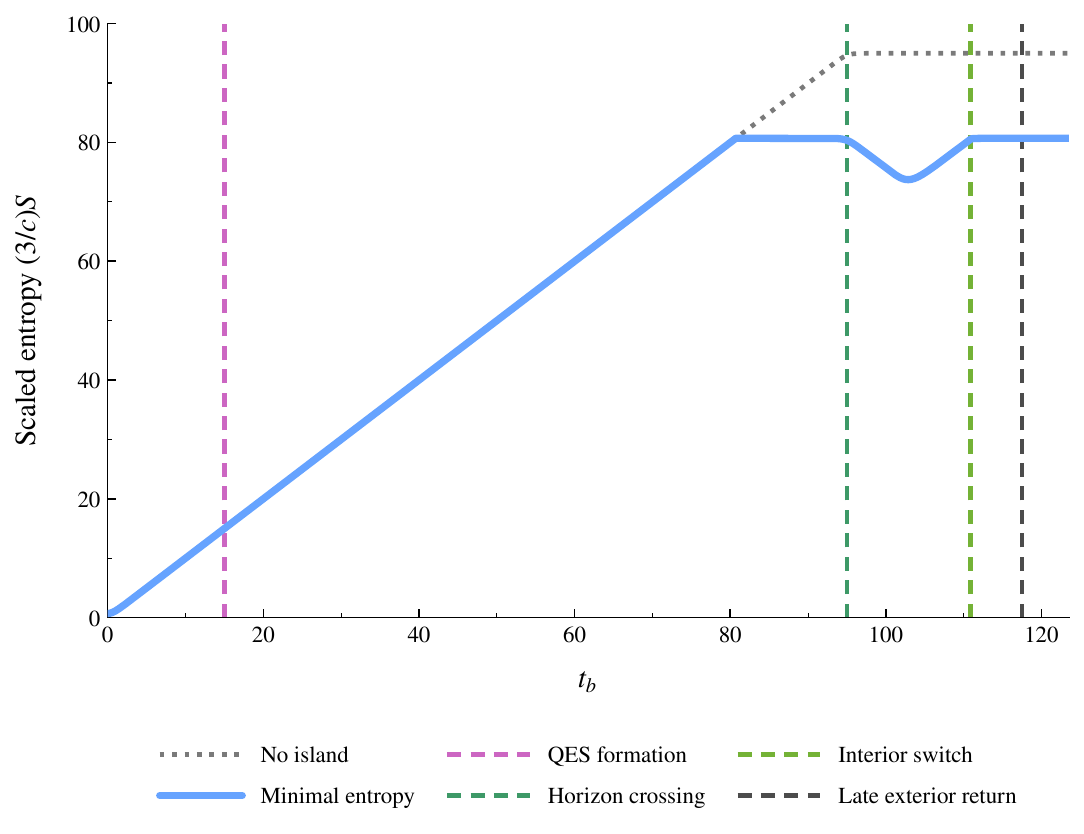}
    \caption{Entropy branches for $b=5$, $L=100$, $\alpha=1/150$,  $6S_0/c=0$, and \mbox{$\kappa\epsilon_b=1$}, with $3S_{\mathrm{BH}}/c=75$.
    \textit{Left}: the no-island and exterior-island entropies, together with the minimum over the three numerically identified interior extrema. \textit{Right}: the minimum entropy envelope compared with the no-island result. The vertical lines mark exterior QES formation, horizon crossing, the  switch between minimizing interior branches, and the late return to an
 exterior island.}
    \label{fig:InteriorBranches}
\end{figure}

Fig. \ref{fig:InteriorBranches} compares the entropy without an island, the entropy when an island is located on the exterior, and the entropy for the interior solution. In the example shown, the interior extrema supply an island branch during the interval in which the near-horizon exterior solution is absent. The minimizing interior branch changes at $t_b\simeq110.919$, where
\begin{equation}
  \frac{3}{c}\bigl(S_F-S_{\mathrm{BH}}\bigr)\simeq5.602.
  \label{eq:int-numerical-switch}
\end{equation}
This is below the ordinary exterior plateau correction
$b+\log2\simeq5.693$. After the Page transition near $t_b\simeq80.693$, the minimum entropy therefore remains island-dominated through the reflection interval, as shown in the right panel of Fig.~\ref{fig:InteriorBranches}.

\section{Backreaction and QES corrections}\label{sec:backreaction}

In the previous sections, we analyzed the quantum extremal surfaces and their contributions to the radiation entropy on the classical JT background. We found that the reflecting wall significantly modifies the QES structure, leading to the disappearance of the exterior branch and the emergence of interior extrema. Since these effects arise from quantum matter correlations, it is interesting to examine whether the corresponding semiclassical backreaction can modify this picture. We do not claim to address these issues in full  generality, but rather focus on some accessible possible implications for the QES structure in our setup.  In this section, we compute the matter stress tensor in the Hartle--Hawking state and determine the resulting correction to the dilaton profile. We then investigate its effect on the QES equations and the generalized entropy.

$\,$

Evaluating the generalized entropy on the classical JT background captures the quantum matter correlations without their backreaction on the dilaton. We now discuss the semiclassical backreaction of the massless Dirac fermions and its effect on the quantum extremal surfaces obtained in previous sections. Following   ~\cite{Pedraza:2021cvx,Pedraza:2021ssc}, we compute the renormalized matter stress tensor in the Hartle–Hawking state and solve the resulting equation for the dilaton. We show that, with the background parameters held fixed, backreaction produces a constant dilaton shift and leaves the QES equations unchanged.

We still consider a massless Dirac field defined throughout the gravitating region and the flat bath,
\begin{equation}
    I_\psi=i\int_{\rm JT+bath} d^2x\sqrt{-g}\bar{\psi}\gamma^\mu\nabla_\mu\psi.
\end{equation}
The action is written for a single complex Dirac field and for several independent species, the corresponding actions are summed, $c$ denotes their total central charge. We introduce null coordinates \mbox{$x^\pm=t_{\rm phys}\pm\sigma_{\rm phys}$}, such that the metric \eqref{eq:MetricNullLor} takes the following form
\begin{equation}\label{eq:NullMetric}
ds^2=-e^{2\rho(x^+, x^-)}dx^+ dx^-,
\end{equation}
where for the JT region the Weyl factor is 
\begin{equation}
    \Omega_\text{JT}^2=e^{-2\rho(x^\pm)}=\frac{\sinh^2\frac{\kappa}{2}(x^+-x^-)}{\kappa^2}.
\end{equation}
In order to correctly glue both regions we place the transparent interface at \mbox{$\sigma_{\rm phys}=-\epsilon$} with \mbox{$\epsilon>0$}. The constant Weyl factor for the bath region \mbox{$\sigma_{\rm phys}\in[0,L_{\rm phys}]$} can be written as 
\begin{equation}
    \Omega_{\text{bath}}=\Omega_\text{JT}(-\epsilon)=\frac{\sinh \kappa\epsilon}{\kappa}.
\end{equation}
Therefore, the two background metrics agree with each other at the transparent interface, 
\begin{equation}
    ds^2_\text{JT}\big|_{\epsilon}=ds^2_\text{bath}\big|_{\sigma=0}.
\end{equation}
The action for fermions can be conveniently rewritten after the Weyl rescaling in terms of a field $\chi_\pm$, which is defined as
\begin{equation}
    \psi=e^{-\frac{\rho}{2}}
    \begin{pmatrix}
        \chi_+\\
        \chi_-
    \end{pmatrix}.
\end{equation}
Then in the null coordinates the action becomes
\begin{equation}\label{eq:FermActionChi}
    I_\chi=-i\int dx^+ dx^-\left(\chi_+^\dagger\partial_-\chi_+ +\chi_-^\dagger\partial_+\chi_-\right),
\end{equation}
where two chiralities propagate independently in the bulk,
\begin{equation}
    \partial_-\chi_+=0,\quad \partial_+\chi_-=0,
\end{equation}
and are related by the reflecting boundary condition at the wall. We impose the proper matching conditions on $\chi$ at the interface,
\begin{equation}
    \chi_{\text{JT},+}(t,-\epsilon)=\chi_{\text{bath},+}(t,0),\quad \chi_{\text{JT},-}(t,-\epsilon)=\chi_{\rm bath,-}(t,0),
\end{equation}
so that the chiralities are not mixed. Since the Weyl factors agree there, they also imply
$\psi_{\rm JT}(t,-\epsilon)=\psi_{\rm bath}(t,0)$.

Given that classical JT gravity is now coupled to matter, we find that the backreaction alters the gravitational equation of motion \eqref{eq:StressTensorDilaton} as
\begin{equation}\label{eq:StressTensorPhi}
    T_{\mu\nu}^\Phi+ \langle T_{\mu\nu}^\chi\rangle = 0.
\end{equation}
Although the field redefinition removes the conformal factor from the classical action, the quantum measure is not Weyl invariant. The renormalized matter stress tensor therefore contains a conformal anomaly. In the bulk, the trace anomaly is
\begin{equation}
    \langle {T^{\chi,\,\mu}}_\mu\rangle = \frac{c}{24\pi}R.
\end{equation}
For the metric \eqref{eq:NullMetric} the Ricci scalar is
\begin{equation}
   R=8e^{-2\rho}\partial_+\partial_-\rho,
\end{equation}
and therefore the mixed components of the stress tensor become
\begin{equation}\label{eq:StressTensor+-}
    \langle T_{+-}^\chi\rangle=\langle T_{-+}^\chi\rangle=-\frac{c}{12\pi}\partial_+\partial_-\rho(x^\pm).
\end{equation}
Using the conservation of the stress tensor, \mbox{$\nabla_\mu {T^\mu}_\nu=0$}, we can calculate the remaining components of the stress tensor in curved space,
\begin{equation}\label{eq:StressTensor++}
    \langle T_{\pm\pm}^\chi(x^\pm)\rangle=-\frac{c}{12\pi}t_\pm(x^\pm)-\frac{c}{12\pi}\bigg[\left(\partial_\pm\rho\right)^2-\partial^2_\pm\rho\bigg].
\end{equation}
The functions $t_\pm(x^\pm)$ encode the state dependence. They are related to the expectation values of the normal ordered chiral stress tensor for some state $|\Psi\rangle$,
\begin{equation}
    \langle \Psi|: T_{\pm\pm}^\chi(x^\pm):|\Psi\rangle= -\frac{c}{12\pi} t_\pm(x^\pm).
\end{equation}
To determine these functions, we use Euclidean chiral correlators and then analytically continue the resulting stress tensor to Lorentzian signature. The calculation is identical for the two chiralities, so we temporarily suppress the $\pm$ indices. For one fermion species in the vacuum defined by a conformal coordinate $U$, the chiral two-point function is
\begin{equation}
    \langle 0_U|\chi(U_1)\chi^\dagger(U_2)|0_U\rangle=\frac{1}{2\pi(U_1-U_2)}.
\end{equation}
Since the chiral fermion has conformal weight $1/2$, the correlator in the $x$ coordinate is
\begin{equation}\label{eq:ChiCorr}
    G(x,x')=\langle 0_U|\chi(x)\chi^\dagger(x')|0_U\rangle=\frac{1}{2\pi}\frac{\sqrt{U'(x)U'(x')}}{U(x)-U(x')}.
\end{equation}
By choosing the points \mbox{$x=X+\delta/2$} and \mbox{$x'=X-\delta/2$} symmetrically and expanding for small values of $\delta$, we can rewrite the right-hand side of \eqref{eq:ChiCorr} as
\begin{equation}\label{eq:DeltaExpand}
    \frac{\sqrt{U'(x)U'(x')}}{U(x)-U(x')}=\frac{1}{\delta}+\frac{\delta}{12}\{U,X\}+\mathcal{O}(\delta^3),
\end{equation}
where $\{U,X\}$ is the Schwarzian derivative. By substituting the derivatives of $G(x,x')$ into the expression for the stress tensor,
\begin{equation}
    \begin{aligned}
        T^\chi_{xx}(X)&=\frac{1}{2}\lim_{\delta\to0}\bigg[\chi^\dagger(x)\partial_{x'}\chi(x')-\partial_{x}\chi^\dagger(x)\chi(x')\bigg]_{\rm ren},\\
        \langle 0_U|:T^\chi_{xx}(X):|0_U\rangle&= -\frac{1}{2}\lim_{\delta\to 0}\bigg[\partial_x G(x,x')-\partial_{x'}G(x,x')\bigg]_{\rm ren},
    \end{aligned}
\end{equation}
we obtain that the singular part of \eqref{eq:DeltaExpand} cancels out and $t_{xx}$ is simply expressed in terms of the Schwarzian
\begin{equation}
    \langle 0_U|:T^\chi_{xx}(X):|0_U\rangle=-\frac{c}{24\pi}\{U,X\},\quad t_{xx}(X)=\frac{1}{2}\{U,X\},
\end{equation}
where we restored central charge $c$ for the total stress tensor. 

$\,$

We now apply this result to the BCFT state used in the previous sections. Consider the vacuum on the upper half-plane with respect to the $z$-coordinate. The upper half-plane is mapped to a disk $\abs{w}< R$ with $R=e^L$ by the transformation \eqref{eq:MapToDisk}, which makes the corresponding Schwarzian to be zero,
\begin{equation}
    t_{w}=\frac{1}{2}\{z,w\}=0.
\end{equation}
For the thermal coordinates that are related to the disk by the exponential map
$w^+ = e^{\kappa x^+}$ and $w^- = e^{-\kappa x^-}$, we find that 
\begin{equation}
    t_\pm=\frac{1}{2}\{w,x^\pm\}=-\frac{\kappa^2}{4},
\end{equation}
which is exactly the Hartle–Hawking vacuum \cite{Pedraza:2021cvx}. We note that in the Euclidean signature, fermions must satisfy the following condition
\begin{equation}
    \psi(\tau+\beta)=-\psi(\tau).
\end{equation}
For the BCFT vacuum $|0\rangle$ on the upper-half plane the reflecting boundary only appears in the mixed correlator,
\begin{equation}
    \langle 0|\chi_+(z)\chi^\dagger_-(\bar{z}')|0\rangle=\frac{1}{2\pi(z -\bar{z}')},
\end{equation}
while the chiral components $T_{++}$ and $T_{--}$ are constructed from the
corresponding same-chirality correlators. This allows us to conclude that the finite size of the bath does not introduce any additional dependence on the length scale $L$ in the stress-energy tensor, and also that the vacuum on the upper half-plane corresponds to the choice of the Hartle-Hawking state.

Next, we calculate the components of the stress tensor in two regions: the curved JT region and the flat bath. Substituting the found values of $t_\pm$ and the corresponding Weyl factors $\rho(x^\pm)$ into equations \eqref{eq:StressTensor+-} and \eqref{eq:StressTensor++}, we obtain the expectation values of the components of the stress-energy tensor with respect to $\chi$ in the Hartle-Hawking state. In the bath region, $\rho$ is constant, and therefore we have
\begin{equation}
    \langle 0|T_{\pm\pm}^\chi|0\rangle_{\rm bath}=\frac{c\kappa^2}{48\pi}, \qquad
    \langle 0|T_{+-}^\chi|0\rangle_{\rm bath}=0.
\end{equation}
For the JT black hole region the state-dependent and geometric terms in the chiral components cancel,
giving
\begin{equation}
    \langle 0|T_{\pm\pm}^\chi|0\rangle_{\rm JT}=0, \qquad
    \langle 0|T_{+-}^\chi|0\rangle_{\rm JT}=\frac{c\kappa^2}{48\pi}\,\frac{1}{\sinh^2 \kappa\sigma},
\end{equation}
equivalently,
\begin{equation}
    \langle 0|T_{\mu\nu}^\chi|0\rangle_{\rm JT}=-\frac{c}{24\pi}g_{\mu\nu}.
\end{equation}

Substituting this expression into the
semiclassical equation \eqref{eq:StressTensorPhi} and choosing the same stationary homogeneous solution as in \eqref{eq:Dilaton}, we find that in the Hartle-Hawking vacuum, the backreaction causes the dilaton \eqref{eq:Dilaton} to modify by a constant,
\begin{equation}
    \Phi^\text{br}(w,\bar w) = \Phi_h \, \frac{1+w\bar w}{1-w\bar w} +\frac{c G}{3}.
\end{equation}
Returning to the dimensionless coordinate
\mbox{$\sigma=\kappa\sigma_{\rm phys}$}, the symmetric exterior island endpoints are located at \mbox{$\sigma=-a$}. Evaluating the dilaton at these endpoints gives
\begin{equation}
    \begin{aligned}
        \Phi^\text{br}(a)&=\Phi_h\coth(a) +\frac{c G}{3}.
    \end{aligned}
\end{equation}
Due to the backreaction, the dilaton undergoes a constant additive shift, which means that the gravitational contribution to entropy \eqref{eq:GravContrib} also changes by a constant
\begin{equation}
    S_{\rm grav}^{\rm br}=2S_0+\frac{\Phi_h}{2 G}\coth a+\frac{c}{6}.
\end{equation}
Therefore the QES equations \eqref{eq:QESTime} and
\eqref{eq:QESRadial} retain their original form, and the island locations are unaffected by this additive correction. The same constant shift
applies to the two-endpoint interior island saddles discussed above.
Although the relative order of these island saddles remains unchanged, their
entropy crossings with the no-island saddle can shift when the renormalized
parameters are kept fixed.

\section{Conclusion}\label{sec:Conclusion}

In this work we studied the island prescription in Jackiw--Teitelboim gravity coupled to a finite non-gravitating bath bounded by a reflecting conformal wall. Unlike the standard infinite bath setup, the finite extent of the radiation system introduces an additional infrared scale while preserving the transparent interface between the gravitating and non-gravitating regions. For a free massless Dirac CFT, this allows the generalized entropy to be computed analytically, including the complete dependence on the wall through the BCFT image contributions.

We first analyzed the trivial island saddle and obtained an exact expression for the radiation entropy in the presence of the reflecting wall. The finite bath modifies the late time behavior of the entropy by replacing the unbounded linear growth of the infinite bath case with saturation at a value determined by the wall position. 
Consequently, for sufficiently short
baths, the saturated no-island entropy
can remain below the thermodynamic
black hole entropy, avoiding the
usual contradiction associated
with unbounded radiation entropy growth.
More generally, however, the
thermodynamic criterion should be
distinguished from the dominance
condition in the island prescription.
The latter requires comparison of
the complete generalized entropies
and depends on whether the reflected
correlations become important before
or after the Page transition.
This reproduces in the JT framework
the qualitative saturation mechanism
previously observed for black holes
placed inside finite cavities, while
making explicit the additional role
of quantum extremal surfaces.

For larger bath sizes, however, the no-island entropy exceeds the thermodynamic entropy, and the island contribution becomes essential. The existence of a QES and its dominance in the entropy minimization are distinct conditions, governed by parametrically different scales. We derived the generalized entropy for a symmetric island configuration and obtained the corresponding QES equations, both exactly and in the semiclassical near-horizon regime. The resulting approximate analytic solutions show excellent agreement with numerical calculations in the region where this approximation is valid.

Our main result is the appearance of a blinking regime when the extremization problem is restricted to exterior endpoints. The early exterior QES reaches the future horizon when the reflected correlations become significant. An exterior solution reappears only after an additional delay. This behavior is controlled by the interplay between the propagation scale associated with the finite bath and a logarithmic semiclassical scale. The wall dependence enters through the nonlocal correlations contributing to the generalized entropy. In the matter sector considered here, the leading backreaction produces a constant dilaton shift and leaves the QES equations unchanged.

Crucially, the disappearance of the exterior solution does not establish the disappearance of the island itself. The divergent exterior coordinates approach a regular point on the future horizon, motivating an extension of the extremization problem into the black hole interior. Our interior analysis identifies real QES candidates in the interval where the exterior branch is absent. The controlled analytical solution remains close to the horizon, and the numerical example shows that including interior extrema keeps the entropy envelope island dominated after the Page transition, including throughout the exterior blinking interval.

These results suggest that restoring the full island picture, and thereby avoiding the reappearance of the information paradox, requires access to the interior region at the level of the generalized entropy prescription\footnote{The authors used AI tools in preparation of the text and particular coding tasks. Authors reviewed and edited the content as needed and take full responsibility for the content of the publication.}.

\newpage
\appendix
\section{Disappearing of the island in the intermediate time regime}\label{sec:AppBlink}

In this appendix, we show that there is no solution to the QES equations \eqref{eq:QESTime} and \eqref{eq:QESRadial} for the symmetric island configuration within the semiclassical framework
\begin{equation}
    \alpha=cG/3\Phi_h\ll1.
\end{equation}
To begin with, let us rewrite the QES equations in a simpler form
\begin{equation}
    \begin{aligned}
        e^{-(a+b)}\left(1-e^{-2(L-b)}\right)\left(T^{-1}-T\right) +\frac{1}{1+X}+\frac{Y}{1+Y}-\frac{Z}{1+Z} &= 0,\\
        e^{-(a+b)}\left(1-e^{-2(L-b)}\right)\left(T^{-1}+T\right) -\frac{2 e^{-2a}}{\alpha} -\frac{1}{1+X}+\frac{1}{1+Y}-\frac{1}{1+Z} &= 0,
    \end{aligned}
\end{equation}
where 
\begin{equation}
    \begin{aligned}
        X&= \exp[t_a+t_b -(a+b)],\\
        Y&=\exp[t_a+t_b-(2L+a-b)],\\
        Z&=\exp[2t_a-2(L+a)],\\
        T&= \exp[t_a-t_b].
    \end{aligned}
\end{equation}
The intermediate time regime assumes that 
\begin{equation}
    X\gg1,\quad Y\gg 1,\quad Z\ll 1.
\end{equation}
Therefore the equation for the time derivative reduces to 
\begin{equation}\label{eq:AppTimeDer}
    e^{-(a+b)}\left(1-e^{-2(L-b)}\right)\left(T- T^{-1}\right) = 1,
\end{equation}
and the radial equation becomes
\begin{equation}
    e^{-(a+b)}\left(1-e^{-2(L-b)}\right)\left(T^{-1}+T\right)-\frac{2 e^{-2a}}{\alpha}-1=0.
\end{equation}
Combining these equations, we get that 
\begin{equation}
    e^{-a}=\frac{\alpha e^{-b}\left(1-e^{-2(L-b)}\right)}{T}.
\end{equation}
Then the left-hand side of Eq.\eqref{eq:AppTimeDer} is 
\begin{equation}
    \begin{aligned}
        &e^{-(a+b)}\left(1-e^{-2(L-b)}\right)\left(T- T^{-1}\right)\\
        =&\;\alpha e^{-2b}\left(1-e^{-2(L-b)}\right)^2\left(1- T^{-2}\right)\\
        <& \;\alpha e^{-2b}\left(1-e^{-2(L-b)}\right)^2.
    \end{aligned}
\end{equation}
In the semiclassical regime, 
\begin{equation}
    \alpha e^{-2b}\left(1-e^{-2(L-b)}\right)^2 \ll 1,
\end{equation}
which contradicts the required value of $1$ in Eq.\eqref{eq:AppTimeDer} for the time derivative. 

Thus, for intermediate time values, there is no island solution near the horizon in the semiclassical regime.

\section{Unified near-horizon QES equation and critical scales}
\label{sec:AppUnifiedBlinking}

In this appendix we derive the unified near-horizon QES equation used in
Sec.~\ref{sec:UnifiedBlinking}.  We then recover the early, intermediate,
and late time regimes as limiting cases and determine the critical bath size for
the existence of the early branch.

\subsection{Reduction to a single existence condition}

We begin with the near-horizon QES equations obtained in the main text.  It
is convenient to introduce \mbox{$d=L-b $} and \mbox{$F=1-e^{-2d}$}, as well as
\begin{equation}
\label{eq:AppXYZT}
    T=e^{t_a-t_b}, \qquad
    X=e^{t_a+t_b-(a+b)}, \qquad
    Y=e^{t_a+t_b-(2L+a-b)}, \qquad
    Z=e^{2t_a-2(L+a)}.
\end{equation}
In this notation the reduced QES equations take the form
\begin{align}
\label{eq:AppReducedTime}
    0={}& Fe^{-(a+b)}\left(T^{-1}-T\right)
    + \frac{1}{1+X}
    + \frac{Y}{1+Y}
    - \frac{Z}{1+Z},
\\
\label{eq:AppReducedRadial}
    0={}& Fe^{-(a+b)}\left(T^{-1}+T\right)
    - \frac{2e^{-2a}}{\alpha}
    - \frac{1}{1+X}
    + \frac{1}{1+Y}
    - \frac{1}{1+Z}.
\end{align}
Adding Eqs.~\eqref{eq:AppReducedTime} and
\eqref{eq:AppReducedRadial}, one finds
\begin{align}
    0={}& \frac{2Fe^{-(a+b)}}{T}
    - \frac{2e^{-2a}}{\alpha}
    + \left[ \frac{Y}{1+Y} + \frac{1}{1+Y} \right]
    - \left[ \frac{Z}{1+Z} + \frac{1}{1+Z} \right].
\end{align}
The last two brackets are equal to one and cancel each other out. There the remaining equation is
\begin{equation}
\label{eq:AppRadialRelation}
    e^{-a} = \frac{\alpha e^{-b}F}{T}.
\end{equation}
Thus, the position of the island can be expressed in terms of the $T$ function as
\begin{equation}
\label{eq:Appa0}
    a = a_0+\log T, \qquad a_0 \equiv \log\frac{1}{\alpha e^{-b}F},
\end{equation}
and, by the definition of $T$,
\begin{equation}
\label{eq:Appta}
    t_a=t_b+\log T.
\end{equation}
An important consequence is that the combinations $X,Y,Z$ are independent
of $T$.  Indeed,
\begin{equation}
X = \alpha F e^{2t_b-2b},\qquad
Y = e^{-2d}X,\qquad
Z = \alpha F e^{-2L}X.
\label{eq:AppX}
\end{equation}
Let us further define
\begin{equation}
\label{eq:Appp}
    p\equiv\alpha e^{-2b}F^2.
\end{equation}
Equation~\eqref{eq:AppRadialRelation} implies
\begin{equation}
    Fe^{-(a+b)} = \frac{p}{T}.
\end{equation}
Substituting this relation into Eq.~\eqref{eq:AppReducedTime} and introducing
\begin{equation}
\label{eq:AppH}
    \mathcal H(t_b) \equiv \frac{1}{1+X}
    + \frac{Y}{1+Y} - \frac{Z}{1+Z},
\end{equation}
we obtain the QES equation in the form of
\begin{equation}
\label{eq:AppMaster}
    T^{-2} = 1-\frac{\mathcal H(t_b)}{p}.
\end{equation}
Therefore the solution to the QES equation has the following form
\begin{equation}
\label{eq:AppMasterSolution}
    t_a = t_b - \frac12 \log\left( 1-\frac{\mathcal H}{p} \right), \qquad
    a = a_0 - \frac12 \log\left( 1-\frac{\mathcal H}{p} \right).
\end{equation}

In the semiclassical regime relevant for the present analysis,
$Y>Z$, so $\mathcal H>0$.  A real solution of
Eq.~\eqref{eq:AppMaster} therefore exists if and only if
\begin{equation}
\label{eq:AppExistence}
    \mathcal H(t_b)<p.
\end{equation}
This is the unified existence condition quoted in the main text.

\subsection{Early, intermediate, and late limits}

We now show explicitly how the separate solutions discussed in the main text
follow from Eq.~\eqref{eq:AppMaster}.

\paragraph{Early regime.}
For \mbox{$\,X\gg1,\, Y\ll1,\, Z\ll1\,$} we have
\begin{equation}
    \mathcal H = \frac{1}{1+X} + \frac{Y}{1+Y}
    - \frac{Z}{1+Z} \simeq \frac1X+Y-Z.
\end{equation}
Hence
\begin{equation}
    \frac{\mathcal H}{p} \simeq \frac{1}{pX}
    + \frac{Y}{p} - \frac{Z}{p}.
\end{equation}
Using Eqs.~\eqref{eq:AppX}--\eqref{eq:Appp}, we obtain that
\begin{equation}
\label{eq:AppSigmaE}
    \frac{\mathcal H}{p} \simeq
    \frac{ e^{-2t_b} }{ \alpha^2e^{-4b}F^3 }
    + \frac{ e^{2t_b}e^{-2d} }{F}
    - \alpha e^{2t_b-2b}e^{-2d} = \Sigma_E.
\end{equation}
Equation~\eqref{eq:AppMasterSolution} therefore reproduces the early time
solution of the main text.

\paragraph{Intermediate regime.}
For \mbox{$\,X\gg1, \, Y\gg1, \, Z\ll1\,$} the three contributions to $\mathcal H$ behave as
\begin{equation}
    \frac{1}{1+X}\simeq0, \qquad
    \frac{Y}{1+Y}\simeq1, \qquad
    \frac{Z}{1+Z}\simeq0,
\end{equation}
therefore \mbox{$\mathcal H\simeq1$}. On the other hand,
\begin{equation}
    p = \alpha e^{-2b}F^2 \ll1.
\end{equation}
The existence condition \eqref{eq:AppExistence} is therefore violated
\begin{equation}
    \mathcal H\simeq1\gg p.
\end{equation}
This proves the absence of a real symmetric near-horizon QES in the intermediate regime.

\paragraph{Late regime.}
For \mbox{$\,X\gg1,\, Y\gg1,\, Z\gg1\,$} we obtain
\begin{equation}
    \mathcal H = \frac{1}{1+X}
    - \frac{1}{1+Y} + \frac{1}{1+Z} \simeq \frac1X-\frac1Y+\frac1Z.
\end{equation}
It follows that
\begin{equation}
\label{eq:AppSigmaL}
    \frac{\mathcal H}{p} \simeq \frac{1}{pX}
    - \frac{1}{pY} + \frac{1}{pZ} = \Sigma_L,
\end{equation}
which reproduces the late time solution from the main text.

\subsection{Critical bath size}

The early time expression may be written as a quadratic function of
$x=e^{2t_b}$.  From Eq.~\eqref{eq:AppSigmaE},
\begin{equation}
\label{eq:AppAB}
    \Sigma_E = \frac{A}{x} + Bx, \qquad A = \frac{e^{4b}}{\alpha^2F^3}, \qquad B
    = e^{-2d} \left( \frac1F-\alpha e^{-2b} \right).
\end{equation}
The reality condition \mbox{$\Sigma_E<1$} is equivalent to
\begin{equation}
    Bx^2-x+A<0.
\end{equation}
The condition for the positive discriminant is \mbox{$\Delta \equiv 1-4AB >0$}.
Substituting Eq.~\eqref{eq:AppAB}, this gives
\begin{equation}
\label{eq:AppCriticalExact}
    4\, \frac{e^{4b-2d}}{\alpha^2F^3}
    \left(
        \frac1F-\alpha e^{-2b}
    \right) <1.
\end{equation}
For a distant wall,
\begin{equation}
    d\gg1, \qquad F=1+\mathcal O(e^{-2d}),
\end{equation}
and in the semiclassical regime
$\alpha e^{-2b}\ll1$.  Equation~\eqref{eq:AppCriticalExact} becomes
\begin{equation}
    4\frac{e^{4b-2d}}{\alpha^2}<1,
\end{equation}
or
\begin{equation}
\label{eq:Appdc} 
    d>d_c, \qquad d_c = 2b+\log\frac{2}{\alpha}
    +\mathcal O(e^{-2d},\alpha e^{-2b}).
\end{equation}
The separation between the endpoints of the early QES interval can be written as
\begin{equation}
t_+-t_-= \frac12 \log \frac{ 1+\sqrt{\Delta} }{
    1-\sqrt{\Delta} }=\operatorname{arctanh}\sqrt{\Delta}.
\label{eq:AppEarlyWidth}
\end{equation}
Near the critical point we set
\begin{equation}
    d=d_c+\delta d,
    \qquad
    0<\delta d\ll1.
\end{equation}
At leading order in the distant-wall approximation,
\begin{equation}
    \Delta = 1-e^{-2\delta d} \simeq 2\delta d.
\end{equation}
Equation~\eqref{eq:AppEarlyWidth} therefore gives
\begin{equation}
\label{eq:AppSquareRoot} 
    t_+-t_- \simeq \sqrt{2(d-d_c)}.
\end{equation}
The early QES interval thus opens with a square-root law as the bath size is
increased through its critical value.

\newpage
\bibliographystyle{JHEP}
\bibliography{biblio}

\end{document}